\documentclass[
reprint,
superscriptaddress,
amsmath,amssymb,
aps,prl
]{revtex4-2}

\usepackage{amsfonts}

\usepackage{graphicx}
\usepackage{dcolumn}
\usepackage{bm}
\usepackage{hyperref}
\usepackage{physics}
\usepackage{xcolor}
\usepackage{mathrsfs}
\usepackage{mathtools}
\usepackage{comment}

\begin{document}

\title{Universal work statistics in long-range interacting quantum systems}

\author{Andrea Solfanelli}
\email{asolfane@sissa.it}
\affiliation{SISSA, via Bonomea 265, 34136 Trieste, Italy}
\affiliation{INFN, Sezione di Trieste, via Valerio 2, 34127 Trieste, Italy}

\author{Nicol\`o Defenu}
\affiliation{Institut f\"ur Theoretische Physik, ETH Z\"urich, Wolfgang-Pauli-Str. 27 Z\"urich, Switzerland}

\date{\today} 
\begin{abstract} We determine the conditions under which the presence of long-range interactions reduce the energy losses due to defect generation during non-adiabatic evolution, crucial for enhancing the power to efficiency ratio of quantum thermal devices. In order to do so, we investigate the response of long-range systems to diverse external drivings, emphasizing their robustness against dynamic excitation in comparison to generic local systems. This phenomenon is demonstrated through the study of the quantum work statistics, revealing insights into energy transfer efficiency and dynamical quantum criticality. Our results demonstrate the benefits of including a long-range interacting medium for quantum thermodynamics application, highlighting the potential to optimize finite-time quantum thermal cycles. Thanks to the effective dimension approach our findings can be drawn in full generality and, then, specified to different experimentally relevant scenarios.
\end{abstract}

\maketitle

\emph{Introduction.}--Long-range (LR) interacting quantum systems feature coupling energies between microscopic constituents, $J_{i,j}$, that decay as a power law of their distance, $r = |i-j|$, i.e., $J_{i,j}\propto r^{-\alpha}$, with $\alpha > 0$.  These systems have garnered significant interest due to their ability to alter the universal behavior of critical systems, leading to phenomena absent in short-range (SR) interactions.  These include novel dynamical phase transitions\,\cite{Defenu2019dynamical,Halimeh2020quasiparticle}, defect formation\,\cite{Acevedo2014newdynamical,Hwang2015quantum,Defenu2018dynamical,defenu2019universal}, anomalous thermalization\,\cite{Regemortel2016information}, information spreading\,\cite{Tran2020hierarchy,Chen2019finite,Kuwahara2020strictly} and metastable phases\,\cite{Defenu2021metastability,Giachetti2023entanglement}. Moreover, LR systems, due to the robustness against external perturbations\,\cite{xuPhysics2022}, can host robust out-of-equilibrium phases such as discrete Floquet time crystals\,\cite{Russomanno2017Floquet,Pizzi2021higherorder,Giachetti2023fractal,Solfanelli2024stabilization} and high levels of quantum entanglement\,\cite{Solfanelli2023logarithmic}.  
These peculiar properties, combined with the fact that various experimental platforms such as Rydberg atoms, dipolar quantum gases, polar molecules, quantum gases in optical cavities, and trapped ions naturally exhibit LR interactions \cite{Endres2016atom,Labuhn2016tunable,Zeiher2017coherent,hollerith2022realizing,Ritsch2013cold,Mivehvar2021cavity,Blatt2012quantum,Monroe2021programmable}, make LR systems promising platforms for quantum technological applications\,\cite{Defenu2023Longrange}. 

Despite extensive research, the specific advantages of LR interactions in quantum technologies and the conditions under which they outperform nearest-neighbour systems often remain unclear. This question is particularly pressing in  quantum thermodynamics applications, where the need to quantify quantum resources is necessary to establish the technological advantage in real applications\,\cite{smith2022quantum}. This work addresses this gap by identifying conditions under which LR interactions reduce energy losses from defect generation during non-adiabatic evolution. This advantage is crucial in quantum thermodynamics, potentially leading to improve the power outputs and efficiency of the engines.

Our dual objectives are: (1) to explore how LR systems respond to external driving, showing their robustness against dynamic excitation generation and enhanced adiabatic evolution compared to nearest-neighbor systems. These phenomena are captured by the universal properties of the so-called quantum work statistics, emerging as a quantum critical point is approached during the dynamics. Indeed, the work distribution function is known to provide insights into the system's behavior, including information regarding energy transfer efficiency, irreversible processes, steady-state conditions, dynamical quantum criticality, and information scrambling\,\cite{Goold2018therole,Heyl2018dynamical,campisi2017thermodynamics}. 
(2) To demonstrate the advantages of LR interactions in quantum thermodynamics. Indeed, efficient quantum engines face a trade-off between power and efficiency\,\cite{Deffner19Book} due to the second law of thermodynamics (the efficiency of a heat engine must be lower than that of a Carnot engine\,\cite{Fermi1956}) and energy dissipation from dynamic excitations in finite time cycles. 

In a previous study\,\cite{Solfanelli2023quantum}, we suggested that LR couplings may be used to enhance the performance of finite time quantum thermal cycles, specifically in the quasi-static regime of a Quantum Otto cycle featuring an integrable fermionic working substance with LR couplings. Here, we broaden and reinforce our proposal by providing a comprehensive characterization of quantum work statistics in a generic (possibly interacting) LR system undergoing finite-time driving with a velocity ranging between the extremes of a sudden and a slow quench. We establish criteria for LR interactions to reduce non-adiabatic energy losses, focusing on the universal properties of the work distribution by examining both the region near the average value and that of large deviations. To accomplish this in a generic setup, we employ the effective dimension approach to study the universal behavior of systems with LR interactions. Finally, we apply our findings to two experimentally relevant LR interacting systems: the LR quantum Ising chain and the LR Kitaev chain.

\emph{Critical behavior in LR systems: the effective dimension approach:}
In this section we briefly review the main properties characterizing the critical behavior of LR systems and how these may be efficiently understood in terms of the so called effective dimension approach\,\cite{solfanelli2024universality}. 

The presence of non-local interactions $J(r)\propto r^{-\alpha}$ may alter the standard picture provided by the Mermin-Wagner theorem\,\cite{Mermin1966absence} allowing for transition at dimensions smaller than the lower critical one for local systems\,\cite{halperin2018hohenberg}, as observed in various experiments\,\cite{chen2023continuous,feng2023continuous}. Depending on the parameter $\alpha$ three regimes can be identified: (i) for $\alpha\leq\alpha_{\mathrm{mf}}$, where $\alpha_{\mathrm{mf}}$ can be calculated in the mean-field approximation, the mean-field approximation correctly describes the universal behavior; (ii) for $\alpha_{\mathrm{mf}}< \alpha\leq \alpha^*$ the system exhibits peculiar LR critical exponents; (iii) for $\alpha > \alpha^*$ the critical behavior corresponds to the nearest-neighbors ($\alpha\to\infty$) one. 

The effective dimension approach provides an intriguing framework for interpreting these findings. This concept suggests that the critical properties of a LR model in dimension $d$ with a power-law exponent $\alpha$ can be deduced from those of a SR model in an effective fractional dimension $d_{\mathrm{eff}}$. A relation linking this effective dimension to $d$ and $\alpha$ can be determined through general renormalization group arguments leading to \cite{Angelini2014relations}
\begin{align}
    d_{\mathrm{eff}} = \frac{d(p-\eta_{\mathrm{SR}}(d_{\mathrm{eff}}))}{\alpha-d},\label{eq: effective dim}
\end{align}
where $p$ is the power of the low-energy dispersion relation of the SR single particle spectrum (see the SM for additional details \cite{SM}).

A key advantage of the effective dimension approach is its ability to reproduce behavior both within and beyond the mean-field approximation's range by varying a single parameter\,\cite{Angelini2014relations}. Moreover, although Eq. \eqref{eq: effective dim} is not exact for non-Gaussian fixed points, it serves as a highly accurate approximation. It offers an efficient means to estimate critical exponents with minimal error, achieving an accuracy higher than $97\%$ for equilibrium critical exponents when compared to precise numerical estimates, see Ref.\,\cite{solfanelli2024universality}.

\emph{Quantum work statistics:} The study of quantum work statistics examines the dynamics of a quantum system governed by a Hamiltonian  $H(h)$, which depends on an external work parameter $h$. The system starts in the initial state $\rho_i$ and a driving protocol changes the work parameter from $h_i$ to $h_f$. The initial and final Hamiltonians are $H_i = H(h_i) = \sum_n \epsilon_n^i |\epsilon_n^i\rangle\langle\epsilon_n^i|$ and $H_f = H(h_f) = \epsilon_m^f|\epsilon_m^f\rangle\langle\epsilon_m^f|$, respectively. The system evolves unitarily under the influence of the external driving, resulting in the final state $\rho_f = U\rho_i U^\dagger$, with $U = \mathrm{T}\exp\left[-i\int_0^\tau dtH(h_t)\right]$ the unitary evolution operator, where $\mathrm{T}\exp$ denotes the time-ordered exponential. 
During this process, the system exchanges energy with the external driving, manifested as work $W$. Generally, $W$ is a stochastic variable  with a probability density \cite{Talkner2007fluctuation,Goold2018therole}
\begin{align}
    P(W) = \sum_{n,m}p_{n,m}\delta\left(W-(\epsilon_m^f-\epsilon_n^i)\right),
\end{align}
where $p_{n,m}$ represents probabilities associated with energy differences between the energy levels of the initial and final Hamiltonians.  For an initial state incoherent with respect to $H_i$, i.e., $[\rho_i,H_i] = 0$ \cite{Santini2023work}, we have $p_{n,m} = p_n^ip_{n|m}$, where $p_n^i = \mathrm{Tr}[\rho_i|\epsilon_n^i\rangle\langle\epsilon_n^i|]$ is the initial population of the $n$th energy level, and $p_{n|m} = |\langle\epsilon_n^i|U|\epsilon_m^f\rangle|^2$ denotes the transition probability between the $n$th and $m$th energy levels during the unitary evolution\,\cite{Goold2018therole}.

We analyze the probability $P(W)$ through its moment generating function
\begin{align}
    G(s) = \int dW e^{-s W}P(W)
=\mathrm{Tr}\left[U^\dagger e^{-s H_f}Ue^{sH_i}\rho_i\right].\label{eq: def G}
\end{align}
Considering a linear driving protocol, $h(t) = h_i-v t$, within a time interval $t\in [0,\tau]$ and with a quench rate $v = (h_i-h_f)/\tau$. We  may investigate the system's response to different quench velocities $v$. At this scope we focus on the statistics of irreversible work $W_{\mathrm{irr}} = W-\Delta \epsilon_0$, where $\Delta\epsilon_0 = \epsilon_0^f-\epsilon_0^i$ represents the adiabatic work contribution, i.e. the difference between final and initial ground state energies. $W_{\mathrm{irr}}$ accounts for the energy irreversibly dissipated during the evolution due to the dynamic generation of defects in finite-time dynamics. In the following we explore two opposite limits: the sudden quench scenario where $v\to\infty$ ($\tau\to 0$) and the slow quench case where $v\to 0$ ($\tau\to\infty$).

\emph{Sudden quench:} In the sudden quench scenario, assuming the system starts in the ground state \( |\epsilon_0^i\rangle \), Eq.\,\eqref{eq: def G} simplifies to
\begin{align}
    G(s) = \langle\epsilon_0^i|e^{sH_i}e^{-sH_f}|\epsilon_0^i\rangle= e^{-\Delta \epsilon_0s}Z(s)
\end{align}
where $Z(s) = \langle\epsilon_0^i|e^{-(H_f-\epsilon_0^f)}|\epsilon_0^i\rangle$. This expression can be interpreted as the partition function of a $d+1$-dimensional classical system on a film of thickness $s$, with two boundary states $|\epsilon_0^i\rangle$ and a transverse area $L^d$\,\cite{Gambassi2012largedeviations,gambassi2011statistics}. 
Within this interpretation, the cumulant generating function $\mathcal{F} = -\ln G(s)$  represents the free energy per unit temperature of the corresponding classical system\,\cite{Gambassi2012largedeviations}.
As $s$, increases, the free energy density per unit area $f = \mathcal{F}/L^{d}$ decomposes into 
\begin{align}
    f = sf_b+2f_s+f_c(s),\label{eq: free energy density}
\end{align}
where $f_b = \Delta\epsilon_0/L^d$ and $f_s = -(\ln|\langle \epsilon_0^f|\epsilon_0^i\rangle|)/L^d$ can be interpreted as bulk and surface free energy densities in the corresponding classical system in a film geometry\,\cite{Gambassi2012largedeviations,gambassi2011statistics}. The term $f_c(s)$ is a subleading contribution in the large $s$ limit but is significant in critical quenches $h_f = h_c$. Near critical points, $f_c(s)$ embodies the critical Casimir effect and displays the universal scaling form\,\cite{krech1994thecasimir,krech1999fluctuation,gambassi2009thecasimir}: $f_c(s)\approx s^{-d}\Theta(s/\xi)$ for $s\gg a$, where $\xi\gg a$ denotes the correlation length and $a$ represents a microscopic length scale.
The scaling function $\Theta(x)$ is universal and depends solely on the universality class of the bulk classical critical point and the surface universality class \cite{binder2002critical}.

Once the scaling behavior of the characteristic function is established, it reveals universal properties of the work distribution. We examine the intensive irreversible work $w = W_{\mathrm{irr}}/L^d =(W-\Delta\epsilon_0)/L^d$. Our focus lies on the probabilities of large deviation events, where $w$ is significantly smaller than its mean value $\langle w\rangle$, indicating proximity to the adiabatic limit where $w = 0$. The probability of such large fluctuations is expected to decrease exponentially with system size $L^d$, i.e., $P(w)\propto e^{-I(w)L^d}$, where $I(w)$ is the non-negative rate function, vanishing at $w = \langle w\rangle$. The quadratic approximation of $I(w)$ around $w = \langle w\rangle$ reproduces typical Gaussian fluctuations $w-\langle w\rangle \sim 1/\sqrt{N}$ predicted by the central limit theorem \,\cite{touchette2009largedeviation,silva2008statistics,zawadzki2023nongaussian}.

Additionally, we observe that $P(w<0)=0$, implying $I(w<0)= +\infty$, and assuming a bounded spectrum, $P(w>w_M) = 0$, where $w_M$ represents the maximum work that can be introduced into the system, corresponding to a fully filled spectrum. As $L\to\infty$, the rate function is obtained through the Legendre-Fenchel transform\,\cite{touchette2009largedeviation}: $I(w) = -\inf_{s\in\mathbb{R}}\{sw-2f_s-f_c(s)\}$.
\begin{figure*}
    \centering    \includegraphics[width=\linewidth]{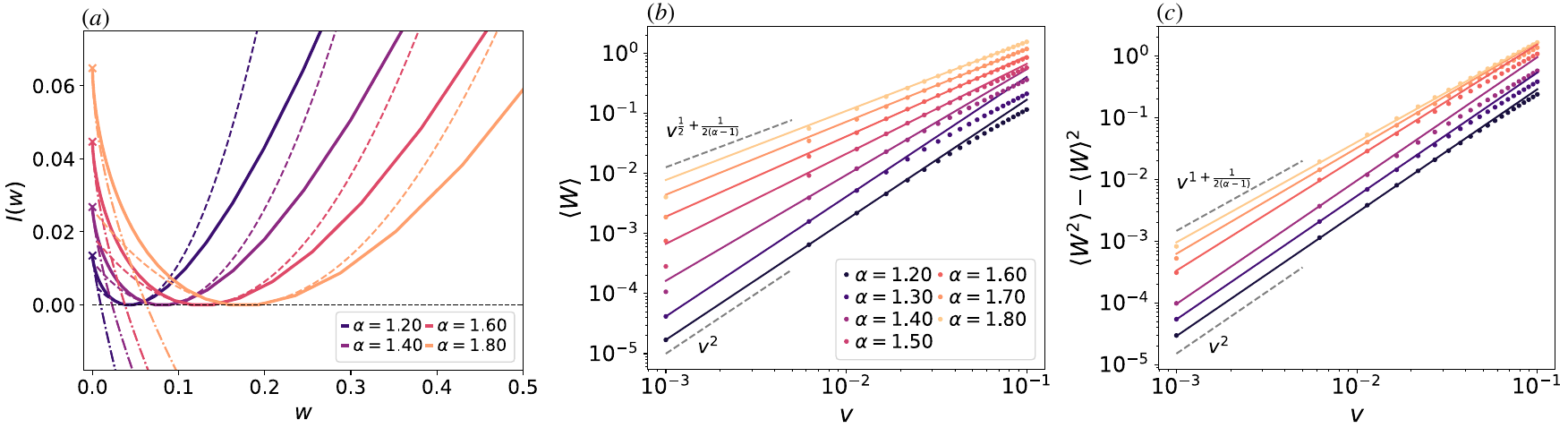}
    \caption{(a) Rate function as a function of the intensive irreversible work $w$ for different values of $\alpha$ following a sudden critical quench of the LR Kitaev chain\,\eqref{eq: LR Kitaev chain H} with $h_i = 21$ and $h_f = 1$. Bold lines depict exact numerical results, dashed lines show the quadratic approximation near the average $w\approx\langle w\rangle$,  and dot-dashed lines illustrate the universal scaling in the large deviation limit $w\ll\langle w \rangle$. (b)-(c) First and second moments of the irreversible work distribution as a function of the driving velocity $v$, for different values of $\alpha$ during slow linear critical driving of LR Kitaev chain\,\eqref{eq: LR Kitaev chain H} with $h_i = 5$ and $h_f = 1$.  Dots represent exact numerical results, while solid lines indicate power law fits of the numerical data. The system size is  $N = 1024$.}
    \label{fig: LD KDQ}
\end{figure*}
In particular, for a critical quench with $\xi\to\infty$, we obtain 
\begin{align}
    I(w)\approx 2f_s-\frac{d+1}{d}\Theta(0)\left(\frac{w}{\Theta(0)}\right)^{d/(d+1)}.\label{eq: I critical}
\end{align}
 LR interactions crucially influence the scaling above. In particular, if we consider a quantum system in $d$ dimensions whose Hamiltonian contains a power-law decaying coupling of the form $J(r)\propto 1/r^\alpha$, the universal properties of the system at criticality are well captured by considering the SR version of the model in a fractal dimension $d_{\mathrm{eff}}$, related to $\alpha$ and $d$ by equation Eq.\,\eqref{eq: effective dim}. Therefore, the out-of-equilibrium work statistics problem in a quantum LR system in $d$  dimensions can be mapped to a classical SR problem at equilibrium in a film geometry in dimension $d_{\mathrm{eff}}+1$. Thus, the rate function for a LR system is obtained by replacing $d$ with $d_\mathrm{eff}$ in Eq.\eqref{eq: I critical}. Typically, $d_{\mathrm{eff}}\geq d$ for any $d<\alpha<\alpha^*$, indicating that in the large deviation region, $I_\mathrm{LR}(w)<I_\mathrm{SR}(w)$, meaning that the probability of having $w\ll \langle w\rangle$ is larger in the LR case. This is a first indication that LR interaction decrease the irreversible work performed by the system.

\emph{Slow quench:} In the slow quench scenario $v\to 0$, it is convenient to write the logarithm of the characteristic function as the series $\ln G(s) = \sum_{n = 1}^{\infty}\frac{s^n}{n!}\kappa_n$, where $\kappa_n$ is the $n$th cumulant of the work distribution. By examining how all moments scale with the (small but finite) driving velocity $v$, one can infer the universal properties of the work distribution in this regime. 

In particular, we consider a time-dependent Hamiltonian of the form $H(\lambda(t))$ which is slowly driven across a quantum critical point at $\lambda=0$. If the critical point can be described by an effective theory with a single quasi-particle branch, the critical energy scales according to the dispersion relation of the low-energy (small $k$) modes $\omega_k = c|k|^z$, where $c$ is a non-zero constant. Under the condition that the total quasi-particles density remains small during the dynamics\,\cite{Polkovnikov2005universal,DeGrandi2010adiabatic}, i.e., $n_k \lesssim 1$ for all $k$, where $n_k$ denotes the occupation number of the $k$th mode, the work performed during the process is given by $W = \Delta\epsilon_0+\sum_k \omega_k n_k$. The probability of having $n_k$ excitations in the $k$th mode is $p_{n_k=0} = 1-p_k$, $p_{n_k=1}\approx p_k$, $p_{n_k>2}\approx 0$. Thus, the logarithm of the characteristic function reads\,\cite{delcampo2018universal,Fei2020work,balducci2023large}
\begin{align}
     \ln G(s) = s\Delta\epsilon_0 + \sum_{k}\ln[1+p_k(e^{-s\omega_k}-1)].\label{eq: G slow quench}
\end{align} 
Then, expanding $\ln G(s)$ in powers of $s$, the cumulants of work are $\kappa_1\approx \Delta\epsilon_0 +\sum_k p_k\omega_k$ and $\kappa_n\approx \sum_k\omega_k^n p_k$\,\cite{Fei2020work}.

For small driving velocity $v\to 0$, the scaling of the work cumulants is determined by the low-energy modes. Therefore, as summarized for completeness in the Supplemental Material\,\cite{SM}, using standard scaling arguments \cite{Polkovnikov2011colloquium}, we obtain
\begin{align}
    \kappa_n - \Delta \epsilon_0 \delta_{n,1} \approx \begin{cases}
        a_n v^{\theta_n}  &\theta_n<2\\
        a_n v^2\ln v      &\theta_n=2\\ 
        a_n v^2           &\theta_n>2
    \end{cases},\label{eq: kn scalings}
\end{align}
with $\theta_n = (d+nz)\nu/(1+\nu z)$.

The universal scalings for a LR interacting system are obtained using the effective dimension approach. The scaling exponent is therefore given by
\begin{align}
    \theta_{n,\alpha} = \frac{(d+nz_{\mathrm{LR}}^d)\nu_{\mathrm{LR}}^d}{1+\nu_\mathrm{LR}^d z_{\mathrm{LR}}^d} \approx  \frac{(d_{\mathrm{eff}}+nz_{\mathrm{SR}}^{d_{\mathrm{eff}}})\nu_{\mathrm{SR}}^{d_{\mathrm{eff}}}}{1+\nu_{\mathrm{SR}}^{d_{\mathrm{eff}}} z_{\mathrm{SR}}^{d_{\mathrm{eff}}}}.\label{eq: scaling effective dim}
\end{align}
The approximate sign accounts for small corrections due to the anomalous dimension in the frequency dependence of the low-energy propagator of the quantum LR theory\,\cite{solfanelli2024universality}. Eq.\,\eqref{eq: scaling effective dim} readily provides a means to estimate LR advantage. If a value $\alpha_{\rm adv}$ exists such that
\begin{align}
\theta_{n,\alpha}>\frac{(d+nz_{\mathrm{SR}}^d)\nu_{\mathrm{SR}}^d}{1+\nu_{\mathrm{SR}}^d},\label{eq: LR advantage}
\end{align}
then, the addition of LR couplings improves the performance of the slow quench work protocol by reducing the irreversible work dissipated during the evolution.

\emph{An analytic example:} We now provide few concrete examples where the applicability of the advantage principle in Eq.\,\eqref{eq: LR advantage} can be verified explicitly. The first one is in an analytically solvable model corresponding to an $N$ sites $p$-wave BCS superconductor with time dependent chemical potential $h$. Assuming periodic boundary conditions, the Hamiltonian of the system is given by
\begin{align}
H(t) = &-\sum_{j=1}^N\sum_{r=1}^{N/2-1}\left[t_r \hat{c}_{j+r}^\dagger \hat{c}_j+\Delta_r\hat{c}_{j+r}^\dagger \hat{c}_j^\dagger +h.c.\right]\notag\\
&-h(t)\sum_{j=1}^N\left[1-2\hat{c}_j^\dagger \hat{c}_j\right].\label{eq: LR Kitaev chain H}
\end{align}
Here, $\hat{c}_j^\dagger$ and $\hat{c}_j$ are fermionic creation and annihilation operators at site $j$, and $t_r, \Delta_r\propto r^{-\alpha}$ are the hopping and pairing amplitudes, respectively. This model is also known as the LR Kitaev chain \cite{Defenu2019dynamical}. The translational invariance and the quadratic nature of the Hamiltonian\,\eqref{eq: LR Kitaev chain H} allows its exact diagonalization in Fourier space via a Bogoliubov transformation (see the SM for more details). The single particle spectrum of the Bogoliubov quasi-particles reads $\omega_k(h) = 2\sqrt{(h-t_k)^2+\Delta_k^2}$. 
Accordingly, the system has two quantum critical points for values of the order parameter $h_c=1,-1+2^{\alpha-1}$ such that $\omega_{k=0,\pi}(h=h_c)=0$.

In the fast quench regime, the cumulant generating function is given by
\begin{align}
    \ln G(s) = -s\Delta \epsilon_0 t+\sum_{k>0}\ln\left[\frac{1+\tan^2\delta\phi_ke^{-2s\omega_{k,2}}}{1+\tan^2\delta\phi_k}\right]
\end{align}
where $\delta \phi_k = \phi_k^f - \phi_k^i$ is the difference between the Bogoliubov angles diagonalizing the final and initial Hamiltonians. In the weak LR regime $1<\alpha<2$ the quasiparticle spectrum becomes continuous in the thermodynamic limit, allowing the sums over Fourier modes to be replaced by integrals. Analyzing the integral in the $s\to\infty$ limit, for a critical quench $h_2 = 1$, the critical free energy density $f_c(s)$ scales as $f_c(s) \propto s^{-\frac{1}{\alpha-1}}$ \cite{SM}.
This matches the effective dimension prediction, confirming $d_\mathrm{eff} = 1/(\alpha-1)$, in agreement with Eq.\eqref{eq: effective dim} with $p=1$. Finally, by explicitly carrying out the Legendre transform, we obtain the rate function
\begin{align}
    I(w) = 2f_{s,\alpha}-\alpha K_\alpha\left(\frac{w}{K_\alpha}\right)^{1/\alpha}\label{eq: I_KLR}
\end{align}
where $K_\alpha$ is an $\alpha$-dependent factor, with its explicit expression provided in the Supplemental Material\,\cite{SM}

Figure \ref{fig: LD KDQ}(a) illustrates  $I(w)$ as a function of $w$ for various values of $\alpha$. As predicted by the general theory discussed in previous sections, $I(w)$ is quadratic for $w\simeq \langle w\rangle$ (dashed lines represents the Gaussian approximation), while it follows the power-law behavior described in Eq. \eqref{eq: I_KLR} for $w\ll \langle w\rangle$. Additionally, as $\alpha$ decreases, indicating stronger LR interactions, $I(w<\langle w\rangle)$ decreases, whereas $I(w>\langle w\rangle)$ increases. This shows that systems with more pronounced LR interactions are more likely to exhibit smaller irreversible work, implying reduced nonadiabatic energy losses during critical dynamics.

For slow driving protocols, Eq.\,\eqref{eq: G slow quench} becomes exact for the model in Eq.\,\eqref{eq: LR Kitaev chain H}. In the $v\to0$ limit, the non-adiabatic transition probabilities can be approximated by $p_k \simeq \exp\left(-\pi\Delta_k^2/v\right)$. Substituting this into the integral and focusing on the contributions from low-energy modes, the resulting expression for $f(s)$ reveals the scaling behavior of the cumulants. Specifically, the $n$th cumulant scales as $\kappa_n \propto v^{\frac{1}{2(\alpha-1)} + \frac{n}{2}}$ as long as $(5-n)/(4-n)<\alpha<2$. This scaling saturates to the SR value $\kappa_n \propto v^{\frac{1+n}{2}}$ for $\alpha\geq 2$, and to $\kappa_n \propto v^{2}$ for $1<\alpha\leq (5-n)/(4-n)$. 
These results are consistent with the effective dimension prediction in Eqs.\,\eqref{eq: kn scalings} and\,\eqref{eq: scaling effective dim}, again with $d_\mathrm{eff} = 1/(\alpha-1)$. Figure\,\ref{fig: LD KDQ} shows the exact numerics for the first (panel (b)) and second (panel (c)) cumulants of the work distribution plotted against the driving velocity. There is good agreement with the predicted scaling, with small corrections arising from finite-size and finite-velocity effects\,\cite{delcampo2018universal}. Most significantly, the LR advantage condition identified by Eq.\,\eqref{eq: LR advantage} is satisfied for this example for any $1<\alpha<2$.
 
\begin{figure}
    \centering    \includegraphics[width=1\linewidth]{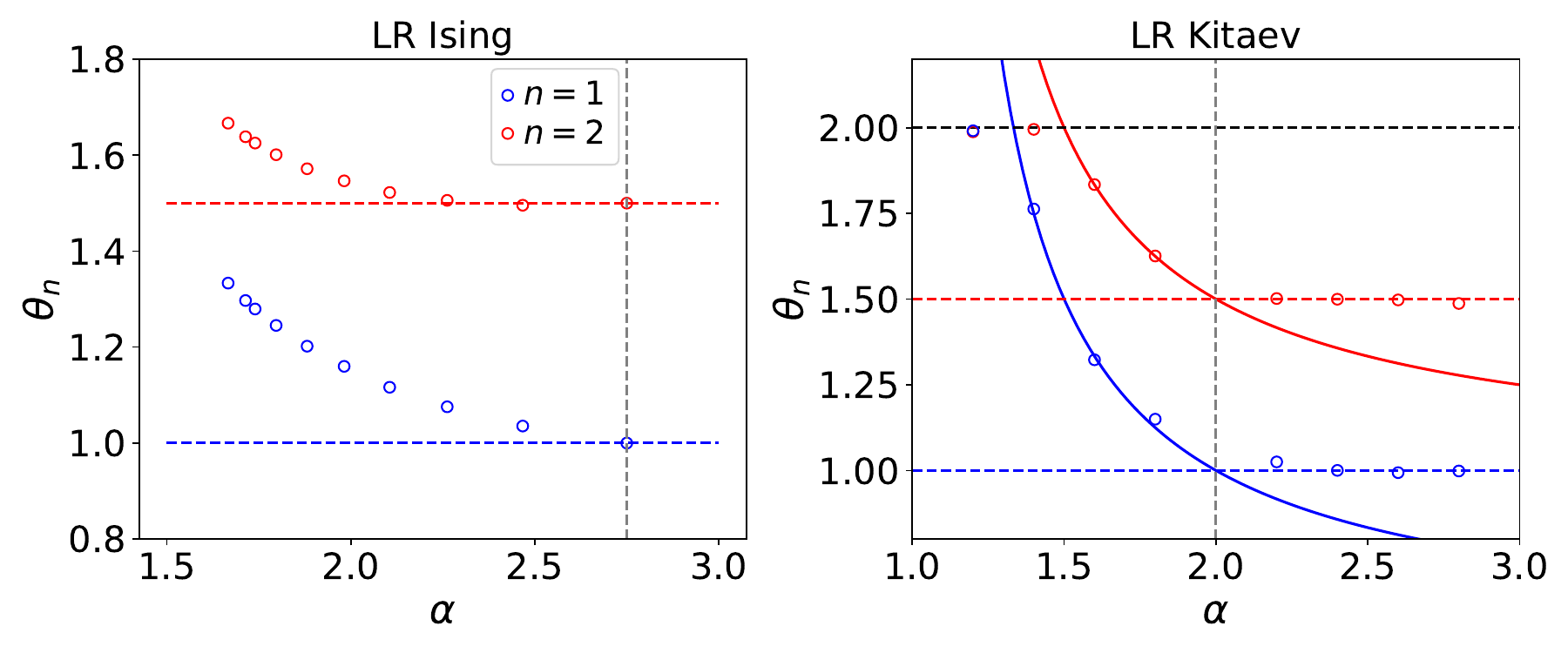}
    \caption{Scaling exponents for the first two cumulants, $\kappa_1$ (blue) and $\kappa_2$ (red) , of the irreversible work statistics with respect to the driving velocity $v$, plotted as a function of $\alpha$The left panel shows results for a slow quench of the LR Ising model, while the right panel displays results for the LR Kitaev chain. In the LR Ising case, dots represent numerical data obtained by combining the effective dimension prediction with precise numerical estimates of the SR Ising critical exponents in the effective fractal dimension $d_\mathrm{eff}$. For the LR Kitaev chain, dots indicate the exponents obtained by fitting the exact numerical data for the work cumulants as a function of $v$ with a power law function. with a power-law function. Horizontal dashed lines represent the SR values of the scaling exponents, and the gray vertical dashed line marks the value of $\alpha=\alpha^*$, above which the SR results apply.}
    \label{fig: theta alpha}
\end{figure}

\emph{Application to experiments:}  Most LR experimental platforms can be described by a spin Hamiltonian\,\cite{Defenu2023Longrange}. Therefore, we explicitly consider the quantum Ising chain with LR interactions
\begin{align}
    H(t) = -\sum_{i<j}J_{i,j}\hat{\sigma}_i^{z}\hat{\sigma}_j^{z}-h(t)\sum_{i}\hat{\sigma}_i^x,
\end{align}
where the $\hat{\sigma}_j^{a}$, $a = x,y,z$ are the Pauli matrices defined at site index $j$. In this case, the scaling exponents for the work cumulants $\theta_n$ in the slow quench scenario can be approximated by inserting precise numerical estimates for the critical exponents of the Ising model in $d_\mathrm{eff}$ dimension\,\cite{elshowk2014conformal} into the effective dimension prediction of Eq.\,\eqref{eq: scaling effective dim}. The resulting $\theta_n$ for the first two cumulants $n = 1,2$ are plotted as a function of $\alpha$ in Fig. \ref{fig: theta alpha}(b) and compared with their corresponding SR values (horizontal dashed lines). Notably, in this interacting example, the first two cumulants of the work statistics satisfy the LR advantage condition \eqref{eq: LR advantage} as long as $\alpha<\alpha^*$, with $\alpha^*\approx 3-\eta_\mathrm{SR}(d_\mathrm{eff} = 1)=2.75$. 

\emph{Conclusion:} In this study, we have explored the peculiar properties of LR interacting quantum systems, particularly their robustness against dynamic excitation generation during non-adiabatic dynamics compared to SR systems. By examining the universal behavior of quantum work statistics, we have identified conditions under which LR interactions reduce the energy losses during a non-adiabatic evolution. This characteristic is crucial for improving the efficiency and power output of quantum engines, particularly in finite-time quantum thermal cycles.

In particular in the sudden quench scenario we showed that the work statistics in a quantum LR system can be mapped to a classical SR problem on a slab geometry in a higher effective dimension $d_\mathrm{eff}+1$. This mapping reveals that the probability of observing reduced irreversible work, much smaller than the average, in LR systems is higher than in SR systems, thus confirming that LR interactions effectively decrease non-adiabatic energy losses during rapid quenches. 

Moreother, for slow driving protocols, we derived the scaling behavior of the cumulants of the work distribution. The effective dimension approach accurately predicts these scalings, showing that LR systems exhibit a distinct advantage in reducing non-adiabatic excitations over a broad range of interaction exponents $\alpha$. Specifically, we identified a critical range of $\alpha$ values where the scaling exponents $\theta_n$ exceed those of SR systems, highlighting the LR advantage.

Our general findings were substantiated through two prototypical examples: the LR Kitaev chain and the LR quantum Ising chain. In the Kitaev chain, we analytically computed the work statistics, demonstrating excellent agreement with our effective dimension predictions. For the Ising chain, numerical estimates of critical exponents validated the LR advantage condition across the significant range of $\alpha\in [d/2,\alpha^*]$ values, where LR interactions actually play a crucial role affecting the system universal properties.

Overall, our comprehensive characterization of quantum work statistics in LR systems provides a robust framework for understanding their enhanced performance in quantum thermodynamic cycles. These insights pave the way for developing more efficient quantum thermal engines and leveraging the unique properties of LR interactions.

\emph{Acknowledgements:}
 A.S. acknowledges useful discussions with Andrea Gambassi while working on a project related to the present work. N.D. acknowledges funding by the Swiss National Science Foundation (SNSF) under project funding ID: 200021 207537 and by the Deutsche Forschungsgemeinschaft (DFG, German Research Foundation) under Germany’s Excellence Strategy EXC2181/1-390900948 (the Heidelberg STRUCTURES Excellence Cluster).
 

\clearpage
\setcounter{equation}{0}
\renewcommand{\theequation}{S.\arabic{equation}}
\setcounter{figure}{0}
\renewcommand{\thefigure}{S.\arabic{figure}}

\begin{widetext}
    
\begin{center}
     \textbf{\Large Supplemental Material}
\end{center}
\section{The effective dimension approach}
To introduce and justify the effective dimension approach, we consider two prototypical classes of LR interacting models: LR quadratic fermionic models and LR $\mathrm{O}(N)$-symmetric spin models, including the LR Ising model as a special case with $N = 1$.

The universality of these models is well captured by the field theory described by the following action:
\begin{align}
S = \int \frac{d^dk}{(2\pi)^d}\omega(k)|\boldsymbol{\varphi}(k)|^2+u\int d^dx|\boldsymbol{\varphi}(x)|^4,\label{eq: action}
\end{align}
where $\omega(k) = a_\alpha k^{\alpha-d} + a_p k^p + r$, whith $p$ the leading analytic contribution in the dispersion relation, and $\boldsymbol{\varphi}$ is an $N$-component bosonic field. In particular, in the examples of the following sections we will have $p = 2$ for the LR quantum Ising chain, while $p = 1$ in the case of the LR Kitaev chain.

At a Gaussian level $(u = 0)$, LR interactions become relevant for $\alpha<d+p$, so that, $\alpha^* = d+p$. The corresponding length dimension of the field is given by
\begin{align}
    [\varphi]\sim \begin{cases}
        L^{-(2d-\alpha)/2} &\mathrm{for}\quad\alpha<d+p\\
        L^{-(d-p)/2} &\mathrm{for}\quad\alpha>d+p
    \end{cases}.
\end{align}
It follows that, at criticality, for $\alpha<d+p$, we have
\begin{align}
    \langle\varphi(x)\varphi(0)\rangle\sim \frac{1}{x^{2d-\alpha}} = \frac{1}{x^{d-p+\eta(\alpha)}},
\end{align}
where $\eta(\alpha) = d+p-\alpha$ is an anomalous dimension only with respect to the usual definition, but it is already present at the Gaussian level. Thus, it can be thought of as a measure of how the decay of correlations differs from the SR one in the mean-field case.

Let us now introduce interactions into our framework as a perturbation of the Gaussian theory. Considering the quartic term in the action \eqref{eq: action}, we have $[\varphi^4]\sim L^{-2(2d-\alpha)}$ for $\alpha<d+p$, so that the perturbation is irrelevant as long as $\alpha < 3d/2$. Accordingly, in this regime, mean-field results are exact, and therefore, for $\mathrm{O}(N)$ models, $\alpha_{\mathrm{mf}} = 3d/2$. To probe values of $\alpha$ outside the mean-field region, one can use perturbative RG around the $d = 4, \alpha = 6$ Gaussian fixed point, expanding in terms of $\epsilon = 3d-2\alpha$.

This problem was first addressed by Sak in the seminal paper  \cite{Sak1973recursion}. One of the main findings of this study is the observation that the $\propto k^{\alpha-d}|\boldsymbol{\varphi}(k)|^2$ term in the action\,\eqref{eq: action} does not acquire an anomalous scaling. Intuitively, this may be traced back to the fact that the perturbative expansion can only generate integer powers of $k^p$, not affecting the non-analytic $\propto k^{\alpha-d}$ behavior.
As a consequence, even in the presence of interactions, the scaling dimension of the LR kinetic term in the Hamiltonian is given by $\Delta_\alpha = 2\Delta_{\varphi}+\alpha-d$. However, by definition, in the SR limit $\Delta_{\varphi} = (d - p + \eta_{\mathrm{SR}})/2$, so we have that
$\Delta_\alpha = \eta_{\mathrm{SR}}-p+\alpha$. Therefore, we find that the LR perturbation is relevant for  $\alpha<\alpha^* = d+p-\eta_{\mathrm{SR}}$. 

To justify Eq.,\eqref{eq: effective dim}, we follow the procedure introduced in Refs. \cite{Banos2012correspondence,Angelini2014relations}. We consider the general scaling form of the singular part of the free energy density for a $LR$ system in $d$ dimensions and a $SR$ system in $d_{\mathrm{eff}}$ dimensions and equate them, leading to
\begin{align}
    f_s &= \frac{1}{V}\Phi_{\mathrm{LR}}(V^{y_{\tau}^{\mathrm{LR}}/d}\tau,V^{y_{h}^{\mathrm{LR}}/d}h,V^{y_{u}^{\mathrm{LR}}/d}u)\notag\\
    &=\frac{1}{V}\Phi_{\mathrm{SR}}(V^{y_{\tau}^{\mathrm{SR}}/d_{\mathrm{eff}}}\tau,L^{y_{h}^{\mathrm{SR}}/d_{\mathrm{eff}}}h,V^{y_{u}^{\mathrm{SR}}/d_{\mathrm{eff}}}u)
\end{align}
where $V$ is the total number of spins, $\tau$ is the reduced temperature, $h$ is the reduced magnetic field, and $u$ is the coupling of the irrelevant operator that gives the leading corrections. The exponents $y_\tau$ , $y_h$ , $y_u$ are connected to the eigenvalues of the linearized form of the RG transformation around the critical fixed point. Thus, the connection between the exponents is 
\begin{align}
    y^{\mathrm{LR}}(\sigma)/d = y^{\mathrm{SR}}/d_{\mathrm{eff}}.\label{eq: relation y}
\end{align}
Combining this condition with the relations of the 
$y$s with the critical exponents, we obtain \cite{Angelini2014relations}
\begin{align}
    &d\nu_{\mathrm{LR}} = d_{\mathrm{eff}}\nu_{\mathrm{SR}},\quad \frac{p-\eta_{\mathrm{LR}}}{d}= \frac{p-\eta_{\mathrm{SR}}}{d_{\mathrm{eff}}}\notag\\
    &\gamma_{\mathrm{LR}} = \gamma_{\mathrm{SR}},\quad\omega_{\mathrm{LR}}/d = \omega_{\mathrm{SR}}/d_{\mathrm{eff}}.\label{eq: critical exponents relations}
\end{align}
Now, as we have seen from perturbative RG arguments, $\eta_{LR} = \eta(\sigma) = d+p-\alpha$, at least up to corrections of order $\mathcal{O}(\epsilon^3)$. Combining this result with the relation between $\eta_{LR}$ and $\eta_{\mathrm{SR}}$ Eq.\,\eqref{eq: critical exponents relations} leads to the dimensional identity in Eq.\,\eqref{eq: effective dim}.

\section{Universal scaling of the work statistics cumulants through adiabatic perturbation theory}
Within adiabatic perturbation theory, the excitation probability of the $k$th-mode quasiparticle $p_k$ is dominated by (assuming that there is no additional Berry phase) \cite{Polkovnikov2005universal,DeGrandi2010adiabatic,Dziarmaga2010dynamics}
\begin{align}
    p_k\approx \abs{\int_{h_i}^{h_f}dh\langle 1_k(h)|\partial_h|0_k(h)\rangle e^{\frac{i}{v}\int_{h_i}^hdh'\omega_k(h')}}^2\label{eq: pk adiabatic perturbation theory}
\end{align}
where $|n_k(h)\rangle$ denotes the instantaneous energy eigenstate of mode $k$ of $H(h)$ with occupation number $n_k$. In order to remove the quantity $1/v$ in the exponential function in the integral of $p_k$ (Eq.\,\eqref{eq: pk adiabatic perturbation theory}), we introduce two rescaled quantities, $x$ and $y$, defined by
\begin{align}
    h = xv^{1/(1+\nu z)},\quad k = yv^{\nu/(1+\nu z)}.
\end{align}
Also following Ref. \cite{DeGrandi2010adiabatic,Dziarmaga2010dynamics}, we introduce the general scaling argument
\begin{align}
    \omega_k(h) &= |h|^{\nu z}F(k/|h|^\nu),\\
    \langle 1_k(h)|\partial_h|0_k(h)\rangle &= h^{-1} G(k/|h|^\nu),
\end{align}
where $F$ and $G$ are two model-dependent scaling functions satisfying $F (x)\propto x^z$ and $G(x)\propto x^{-1/\nu}$ for $|x|\gg 1$. This is motivated by dimensional considerations and the requirement that the spectrum of the high energy modes should be insensitive to $h$. Thus, assuming a critical quench with $\omega_k(h_f)\approx c|k|^z$, the work cumulants reads 
\begin{align}
    \kappa_n\approx \Delta\epsilon_0\delta_{n,1}+c^nv^{\theta_n}\int\frac{d^dy}{2\pi}|y|^{nz}f(y),
\end{align}
where $\theta_n = (d+nz)\nu/(1+\nu z)$, and we have introduced the function
\begin{align}
    f(y) = \abs{\int_{x_i}^{x_f}\frac{dx}{x}G(y/|x|^\nu)e^{i\int_{x_i}^xdx'|x'|^{\nu z}F(y/|x'|^\nu)}}^2\label{eq: integral}.
\end{align}
Consequently, as long as $\theta_n<2$, the scaling of the work cumulants is dominated by the low energy modes, $f(y)$ is convergent as $v\to 0$, and $\kappa_n-\Delta\epsilon_0\delta_{n,1}\approx a_nv^{\theta_n}$. On the other hand,
when $\theta_n>2$ the integral is not dominated by the low energy modes \cite{Polkovnikov2005universal,Polkovnikov2011colloquium,DeGrandi2010adiabatic} and leading term comes from the high-energy contribution, which can be approximated by the regular analytic adiabatic perturbation theory 
\cite{Polkovnikov2005universal,Polkovnikov2011colloquium,DeGrandi2010adiabatic}, resulting in the quadratic scaling  $\kappa_n-\Delta\epsilon_0\delta_{n,1}\approx a_nv^{2}$. Finally, when $\theta_n = 2$, logarithmic corrections are expected leading to the scaling $\kappa_n-\Delta\epsilon_0\delta_{n,1}\approx a_nv^{2}\ln v$.

\section{The LR Kitaev chain}
\subsection{Diagonalization of the Hamiltonian}
In this Appendix we provide more details on the diagonalization the LR Kitaev chain and the corresponding derivations of for the quantum work statistics in the sudden and slow quench scenarios. Let us start by considering the Hamiltonian of the LR Kitaev chain introduced in the main text
\begin{align}
    H_\mathrm{LRK}(t) = -\sum_{j=1}^N\sum_{r=1}^{N/2-1}\left[t_r \hat{c}_{j+r}^\dagger \hat{c}_j+\Delta_r\hat{c}_{j+r}^\dagger \hat{c}_j^\dagger +h.c.\right]-h(t)\sum_{j=1}^N\left[1-2\hat{c}_j^\dagger \hat{c}_j\right].\label{eq: LR quadratic fermionic Hamiltonian}
\end{align}
Assuming periodic boundary conditions, the Hamiltonian can be diagonalized in Fourier space via the Bogolyubov transformation
\begin{align}
	\hat{c}_k = \cos\frac{\phi_k}{2}\hat{\gamma}_k +\sin\frac{\phi_k}{2}\hat{\gamma}_{-k}^\dagger,
\end{align}
where we have introduced the momentum space fermionic operators
\begin{align}
	\hat{c}_k = \frac{e^{-i\frac{\pi}{4}}}{\sqrt{N}}\sum_{j=1}^N e^{ikj}\hat{c}_j,
\end{align}
where $k = 2\pi n/N$ and $n$ is an integer such that $n = -N/2+1,\dots, N/2$. While the Bogoliubov angles are defined by the conditions $\tan\phi_k = \tilde{\Delta}_k/(h-\tilde{t}_k)$
, where Fourier transforms of the hopping and pairing amplitudes are defined as
\begin{align}
	\tilde{t}_k = \frac{J}{N_{\alpha}}\sum_{r=1}^{N/2-1}\frac{\cos(kr)}{r^{\alpha}},\quad
	\tilde{\Delta}_k =\frac{\Delta}{N_{\alpha}}\sum_{r=1}^{N/2-1}\frac{\cos(kr)}{r^{\alpha}}.\label{eq: hopping,pairing}
\end{align}
Hereafter, we set $J = \Delta = 1$ as the energy scale and work in units of $\hbar = 1$. In terms of the Bogoliubov fermions, the Hamiltonian then takes the diagonal form
\begin{align}
	H = \sum_k \omega_k(h)\left(\hat{\gamma}_k^\dagger\hat{\gamma}_k-1/2\right),
\end{align}
with the quasiparticle spectrum
\begin{align}
	\omega_k(h) = 2\sqrt{(h-\tilde{t}_k)^2+\tilde{\Delta}_k^2}.\label{eq: spectrum}
\end{align}
Since $\omega_k(h)\geq 0$, the ground state corresponds to the Fock space vacuum for the Bogoliubov modes, defined by the condition $\hat{\gamma}_k|\mathrm{gs}\rangle = 0, \forall k$.
\subsection{Taylor expansion of the spectrum around the critical modes}\label{sec SM: Taylor expansion of the spectrum around the critical modes}
The critical properties of the system are encoded in the behavior of the quasiparticle spectrum spectrum\,\eqref{eq: spectrum} at lowest order in $|k-k_c|)$, where $k_c = 0$ at the critical point $h = 1$. This is determined by the Taylor expansion of the Fourier transform of the LR couplings $\tilde{t}_k$ and $\tilde{\Delta}_k$ around $k=0$ at leading order, which crucially depend on the value of $\alpha$ considered. In particular, for $\tilde{t}_k$ we find\,\cite{defenu2019universal}
\begin{align}
\tilde{t}_k &= 1+\sin(\frac{\alpha\pi}{2})\frac{\Gamma(1-\alpha)}{\zeta(\alpha)}k^{\alpha-1}+\mathcal{O}(k^2) &1<\alpha<3\\
\tilde{t}_k &=  1+\frac{2\ln(k)-3}{4\zeta(3)}k^2 + \mathcal{O}(k^3) &\alpha = 3\\
\tilde{t}_k &=  1-\frac{\zeta(\alpha-2)}{2\zeta(\alpha)}k^2 + \mathcal{O}(k^{\alpha-1}) &\alpha > 3
\end{align}
While for $\tilde{\Delta}_k$ we have\,\cite{defenu2019universal}
\begin{align}
 \tilde{\Delta}_k &= \cos(\frac{\alpha\pi}{2})\frac{\Gamma(1-\alpha)}{\zeta(\alpha)} +\mathcal{O}(k) &1<\alpha<2\\
 \tilde{\Delta}_k &= \frac{6}{\pi^2}(1-\ln(k))k +\mathcal{O}(k^3) &\alpha = 2\\
 \tilde{\Delta}_k &= \frac{\zeta(\alpha-1)}{\zeta(\alpha)}k+\mathcal{O}(k^{\alpha-1}) &\alpha > 2
\end{align}
Accordingly, inserting these expansions into the expression for the single particle spectrum $\omega_k = 2\sqrt{(h-\tilde{t}_k)^2+\tilde{\Delta}_k^2}$, for $h = 1$ and $k\approx 0$, we obtain
\begin{align}
    \omega_k &= 2\abs{\Gamma(1-\alpha)/\zeta(\alpha)}|k|^{\alpha-1} +\mathcal{O}(|k|)&1<\alpha<2\\
    \omega_k &= 2\abs{\cos(\alpha\pi/2)\Gamma(1-\alpha)/\zeta(\alpha)}|k|^{\alpha-1} +\mathcal{O}(|k|) &2\leq\alpha<3\\
    \omega_k &= 2\abs{\zeta(\alpha-1)/\zeta(\alpha)}|k| + \mathcal{O}(|k|^{\alpha-1}) &\alpha\geq 3
\end{align}

We also report the Bogoliubov angles $\phi_k = \arctan[\Delta_k/(h-t_k)]/2$ at the critical mode $k\approx 0$, which for $1<\alpha<2$ and $h\neq 1$ reads
\begin{align}
    \phi_k = \frac{\cos(\alpha\pi/2)\Gamma(1-\alpha)}{2|h-1|}|k|^{\alpha-1}+\mathcal{O}(|k|), 
\end{align}
while at the critical point $h = 1$ we have
\begin{align}
    \phi_k = \frac{1}{2}\arctan\left(\frac{1}{\tan(\alpha\pi/2)}\right)+\mathcal{O}(|k|) = \frac{\pi(\alpha-1)}{4}+\mathcal{O}(|k|).
\end{align}
\subsection{Qauntum work statistics in the LR Kitaev chain: fast quench}
Thanks to the quadratic nature of the model, the cumulant generating function of the work statistics, as defined in Eq. \eqref{eq: def G}, can be computed analytically and reads:
\begin{align}
    \ln G(s) = -s\Delta \epsilon_0 t+\sum_{k>0}\ln\left[\frac{1+\tan^2\delta\phi_ke^{-2s\omega_{k,2}}}{1+\tan^2\delta\phi_k}\right]
\end{align}
where $\delta\phi_k = \phi_k^f-\phi_k^i$ s the difference between the Bogoliubov angles diagonalizing the final and initial Hamiltonians, respectively. We can identify the bulk free energy density as
\begin{align}
    f_b = \frac{s\Delta\epsilon_0}{L} = \frac{s}{2L}\sum_{k>0}(\omega_{k,1}-\omega_{k,2}),
\end{align}
the surface free energy density
\begin{align}
    f_s = \frac{1}{L}\sum_{k>0}\ln\left[1+\tan^2\delta\phi_k\right],
\end{align}
and the critical free energy density
\begin{align}
    f_c(s) = -\frac{1}{L}\sum_{k>0}\ln\left[1+\tan^2\delta\phi_ke^{-2s\omega_{k,2}}\right].\label{eq: f_c integral KLR}
\end{align}
In the weak LR regime $(1<\alpha<2)$, the quasiparticle spectrum is continuous in the thermodynamic limit. Accordingly, in the $L\to\infty$ limit, we can replace the sums over the Fourier modes $k$ with integrals. Specifically, we can rewrite the critical free energy density as 
\begin{align}
    f_c(s) = -\int_0^\pi \frac{dk}{2\pi}\ln\left[1+\tan^2\delta\phi_ke^{-2s\omega_{k,2}}\right].
\end{align}
To probe the large deviation region of the work distribution with $w\ll\langle w\rangle$, we need to consider the  $s\to\infty$ limit.In this limit, the integral in Eq. \eqref{eq: f_c integral KLR} is dominated by the low-energy modes close to $k\approx 0$. Therefore, we can use the expansions introduced in the previous section, which for $1<\alpha<2$ lead to
\begin{align}
    f_c(s)\approx-\int_0^\pi \frac{dk}{2\pi}\ln\left[1+\tan^2\left(\frac{\pi(\alpha-1)}{4}\right)e^{-2sC_{\alpha}k^{\alpha-1}}\right],
\end{align}
where $C_\alpha = 2|\Gamma(1-\alpha)/\zeta(\alpha)|$.
Then, making the change of variables $y = 2sC_\alpha k^{\alpha-1}$ we obtain
\begin{align}
    f_c(s)\approx-\frac{1}{s^{1/(\alpha-1)}}\frac{1}{2\pi(2C_\alpha)^{1/(\alpha-1)}}\int_0^{2C_\alpha\pi^{\alpha-1}s}dy y^{\frac{1}{\alpha-1}-1}\ln\left[1+\tan^2\left(\frac{\pi(\alpha-1)}{4}\right)e^{-y}\right]
\end{align}
This integral is convergent and in the $s\to\infty$ limit can be carried out explicitly, leading to
\begin{align}
    f_c(s)\approx \frac{\Gamma\left(\frac{1}{\alpha-1}\right)\mathrm{Li}_{1+\frac{1}{\alpha-1}}\left(-\tan^2\left(\frac{\pi(\alpha-1)}{4}\right)\right)}{2\pi(2C_\alpha)^{1/(\alpha-1)}}s^{-\frac{1}{\alpha-1}},
\end{align}
where $\mathrm{Li}_{x}(z) = \sum_{n=1}^\infty z^n/n^x$ is the polylogarithm function. Comparing this result with the effective dimension prediction $f_c(s)\approx \Theta(0)s^{-d_{\mathrm{eff}}}$, with the identification
\begin{align}
    \Theta(0) = -\frac{\Gamma\left(\frac{1}{\alpha-1}\right)\mathrm{Li}_{1+\frac{1}{\alpha-1}}\left(-\tan^2\left(\frac{\pi(\alpha-1)}{4}\right)\right)}{2\pi(2C_\alpha)^{1/(\alpha-1)}}, \quad d_{\mathrm{eff}} = \frac{1}{\alpha-1},
\end{align}
which agrees with the effective dimension predicted by Eq. \eqref{eq: effective dim} of the main text. Finally, carrying out the Legendre transform explicitly, we obtain the rate function
\begin{align}
    I(w) = 2f_{s,\alpha}-\alpha K_\alpha\left(\frac{w}{K_\alpha}\right)^{\frac{1}{\alpha}} = 2f_{s,\alpha}-\frac{d_{\mathrm{eff}}+1}{d_{\mathrm{eff}}}K_\alpha\left(\frac{w}{K_\alpha}\right)^{\frac{d_{\mathrm{eff}}}{d_{\mathrm{eff}}+1}},
\end{align}
where we have introduced the $\alpha$ dependent factor
\begin{align}
    K_\alpha = d_{\mathrm{eff}}\Theta(0)= -\frac{\Gamma\left(\frac{1}{\alpha-1}\right)\mathrm{Li}_{1+\frac{1}{\alpha-1}}\left(-\tan^2\left(\frac{\pi(\alpha-1)}{4}\right)\right)}{2\pi(\alpha-1)(2C_\alpha)^{1/(\alpha-1)}}.
\end{align}

\subsection{Qauntum work statistics in the LR Kitaev chain: slow quench}
We now consider the opposite regime of an infinitely slow linear quench with $h(t) = h_i-v t$, $t\in[0,(h_i-h_f)/v]$ and $v\to 0$. In this case it is convenient to express the cumulant generating function $G(s)$ in terms of the excitation probabilities $p_k$ of each Fourier mode during the dynamics, leading to
\begin{align}
    \ln G(s) = -s\Delta\epsilon_0+L\int_0^\pi\frac{dk}{2\pi}\ln\left[1+p_k(e^{-2s\omega_{k,2}}-1)\right]
\end{align}
Expanding the logarithm in powers of $p_k (e^{-2s \omega_{k,2}} - 1)$, we get
\begin{align}
    \ln G(s) = -s\Delta\epsilon_0+L\sum_{n=1}^\infty\frac{(-1)^{n+1}}{n}\int_0^\pi\frac{dk}{2\pi}p_k^n\left(e^{-2s\omega_{k,2}}-1\right)^n,\label{eq: lnGs slow}
\end{align}
where the series converges as long as $|p_k(e^{-2s\omega_k^f}-1)|<1$.

In general, the unitary evolution generated by $H_\mathrm{LRK}(h(t))$ is such that it only mixes the states $|0_{k},0_{-k}\rangle$ and $|1_{k},1_{-k}\rangle$, where $|1_{k}\rangle = \hat{c}_k^\dagger|0_{k}\rangle$, for each value of $k$. Consequently, the dynamics of the Kitaev chain can be exactly described by $N$ independent evolution equations, each restricted to the two-dimensional subspace associated with the corresponding $k$-mode\,\cite{Dziarmaga2010dynamics}. These can be cast into a matrix evolution for the Bogoliubov coefficients $u_k$,$v_k$:
\begin{align}
	i\frac{d}{dt}\begin{pmatrix}
		u_k\\
		v_k\end{pmatrix} = 
	\mathcal{H}_k(t)
	\begin{pmatrix}
		u_k\\
		v_k
	\end{pmatrix},\label{eq: unitary evolution}
\end{align} 
with $\mathcal{H}_k = (h(t)-t_k)\sigma_k^{z}+\Delta_k\sigma_k^x$, where $\sigma_k^{(a)}$, $a=x,y,z$ are the sigma Pauli operators. By means of the transformation  $t' = \Delta_k(t_k-h+v t)/v$, this is mapped onto a Landau-Zener-St{\"u}ckelberg-Majorana (LZSM) problem: 
\begin{align}
	i\frac{d}{dt'}\begin{pmatrix}
		u_k\\
		v_k
	\end{pmatrix} = \begin{pmatrix}
		-\Omega_k t' &1\\
		1 & \Omega_k t'
	\end{pmatrix}\begin{pmatrix}
		u_k\\
		v_k
	\end{pmatrix},\label{eq: LZSM}
\end{align}
where $\Omega_k = \delta/\Delta_k^2$. The exact general solution of Eq.~\eqref{eq: LZSM} can be written in terms of Weber (or parabolic cylinder) D-functions $D_\nu(z)$,(see Ref.\,\cite{Dziarmaga2010dynamics}), leading to 
\begin{align}
	&v_k(t') = aD_{-s-1}(-iz)+bD_{-s-1}(iz),\label{eq: exact v}\\
	&u_k(t') = \left(\Omega_kt'-2i\frac{\partial}{\partial t'}\right)v_k(t'),\label{eq: exact u}
\end{align}
with $s = (4i\Omega_k)^{-1}$, $z = \sqrt{\Omega_k}t'e^{i\pi/4}$, and $a,b$ arbitrary complex parameters to be fixed by the initial conditions $u_k(t_i)$, $v_k(t_i)$. Accordingly, the solution of Eq.~\eqref{eq: unitary evolution} reads
\begin{align}
	&|\psi(t)\rangle = \prod_k|\psi_k(t)\rangle,\\
	&|\psi_k(t)\rangle = u_k(t)|0_{k},0_{-k}\rangle+v_k(t)|1_{k},1_{-k}\rangle,
\end{align} 
where $u_k(t) = u_k(t'(t)),v_k(t) = v_k(t'(t))$. 

We can introduce the instantaneous eigenstates of the two-level Hamiltonians $\mathcal{H}_k(t)$ at time $t$, given by
\begin{align}
	|\phi^{\pm}_k(t)\rangle = \bar{u}_k(h(t))|0_{k},0_{-k}\rangle \pm \bar{v}_k(h(t))|1_{k},1_{-k}\rangle,
\end{align}
with $\bar{u}_k(h) = \cos(\phi_k(h)/2)$, $\bar{v}_k(h) = \sin(\phi_k(h)/2)$, where $\phi_k(h) = \arctan(\Delta_k/(h-t_k))$ is the Bogoliubov angle for a chemical potential $h=h(t)$. The non-adiabatic transition probabilities then read
\begin{align}
	p_k(t) &= 1-|\langle \phi_k^{\pm}(t)|\psi_k(t)\rangle|^2\notag\\
	&=1-|\bar{u}_k(h(t))u_k(t)+\bar{v}_k(h(t))v_k(t)|^2.
\end{align}
By inserting the expression for $u_k$, $v_k$ in Eqs. ~\eqref{eq: exact v},~\eqref{eq: exact u}, in the above expression, one obtains an analytical expression for $P_k(t)$\,\cite{VitanovPRA1996}. This exact solution, however, is rather cumbersome. Considering the limit of a slow driving protocol $v\to 0$, with final time $\tau = |h_f-h_i|/v\to\infty$ allows for a simpler description that captures and better grasp the relevant physics involved in the dynamics. In this regime, the first non-trivial correction to $p_k$ takes the celebrated LZSM  
form  
\begin{align}
	p_k \simeq \exp\left(-\frac{\pi\Delta_k^2}{v}\right) + \mathcal{O}(v^2). \label{eq: LZSM formula}
\end{align}
See Ref.\,\cite{DeGrandi2010adiabatic} for its derivation using adiabatic perturbation theory. Although for finite $\Delta_k$ the $O(v^2)$ contributions is leading, as the transition point is crossed, the physics is dominated by the soft modes with small $\Delta_k$. As a consequence, in any relevant thermodynamic quantity, the $\mathcal{O}(v^2)$ contribution in the r.h.s. of\,\eqref{eq: LZSM formula} is negligible with respect to the non-analytic exponential one.

Inserting Eq.\,\eqref{eq: LZSM formula} into the integral in Eq.\,\eqref{eq: lnGs slow}, we obtain
\begin{align}
    f(s) = -\frac{1}{L}(\ln G(s)+s\Delta\epsilon_0)\approx-\sum_{n=1}^\infty\frac{(-1)^{n+1}}{n}\int_0^\pi\frac{dk}{2\pi}e^{-\frac{n\pi\Delta_k^2}{v}}\left(e^{-2s\omega_{k,2}}-1\right)^n.
\end{align}
Due to the exponential decay of $p_k$, only low-energy modes can get excited. Therefore, the integral is dominated by the contributions at small Fourier modes $k$ and we can replace the expressions for $\Delta_k$ and $\omega_{k,2}$ with their expansions around $k = 0$. Consequently, for a critical quench with $h_2 = 1$ and $1<\alpha<2$ we obtain
\begin{align}
    f(s)\approx -\sum_{n=1}^\infty\frac{(-1)^{n+1}}{n}\int_0^\pi\frac{dk}{2\pi}e^{-\frac{n\pi B_\alpha^2}{v}k^{2\alpha-2}}\left(e^{-2sC_\alpha k^{\alpha-1}}-1\right)^n,
\end{align}
where $B_\alpha = \cos(\alpha\pi/2)\Gamma(1-\alpha)/\zeta(\alpha)$. Then, performing the change of variable $y^2 = k^{2\alpha-2}n\pi B_\alpha^2/v$, we have
\begin{align}
    f(s)\approx -\frac{1}{2\pi(\alpha-1)}\left(\frac{v}{\pi B_\alpha}\right)^{\frac{1}{2(\alpha-1)}}\sum_{n=1}^\infty\frac{(-1)^{n+1}}{n}\int_0^{\pi^{\alpha-1}|B_\alpha|\sqrt{\frac{\pi}{v}}}dy y^{\frac{1}{\alpha-1}-1}e^{-y^2 n}\left(e^{-2s\frac{C_\alpha}{B_\alpha} \sqrt{\frac{v}{\pi}}y}-1\right)^n.
\end{align}
Finally, keeping only the leading order contributions as $v\to 0$ we find
\begin{align}
    f(s)&\approx \frac{1}{2\pi(\alpha-1)}\left(\frac{v}{\pi B_\alpha}\right)^{\frac{1}{2(\alpha-1)}}\sum_{n=1}^\infty\frac{1}{n}\int_0^{\infty}dy y^{n+\frac{1}{\alpha-1}-1}e^{-y^2 n}\left(2s\frac{C_\alpha}{B_\alpha}\right)^n \left(\frac{v}{\pi}\right)^{\frac{n}{2}}\\
    &= \sum_{n=1}^\infty f_{n,\alpha}(s)v^{\frac{1}{2(\alpha-1)}+\frac{n}{2}}.
\end{align}
This gives us the scaling of the $n$th cumulant $\kappa_n\propto v^{\frac{1}{2(\alpha-1)}+\frac{n}{2}}$. Also, this scaling is in agreement with the effective dimension prediction with $d_\mathrm{eff} = 1/(\alpha-1)$.
\end{widetext}



\begin{thebibliography}{62}%
\makeatletter
\providecommand \@ifxundefined [1]{%
 \@ifx{#1\undefined}
}%
\providecommand \@ifnum [1]{%
 \ifnum #1\expandafter \@firstoftwo
 \else \expandafter \@secondoftwo
 \fi
}%
\providecommand \@ifx [1]{%
 \ifx #1\expandafter \@firstoftwo
 \else \expandafter \@secondoftwo
 \fi
}%
\providecommand \natexlab [1]{#1}%
\providecommand \enquote  [1]{``#1''}%
\providecommand \bibnamefont  [1]{#1}%
\providecommand \bibfnamefont [1]{#1}%
\providecommand \citenamefont [1]{#1}%
\providecommand \href@noop [0]{\@secondoftwo}%
\providecommand \href [0]{\begingroup \@sanitize@url \@href}%
\providecommand \@href[1]{\@@startlink{#1}\@@href}%
\providecommand \@@href[1]{\endgroup#1\@@endlink}%
\providecommand \@sanitize@url [0]{\catcode `\\12\catcode `\$12\catcode
  `\&12\catcode `\#12\catcode `\^12\catcode `\_12\catcode `\%12\relax}%
\providecommand \@@startlink[1]{}%
\providecommand \@@endlink[0]{}%
\providecommand \url  [0]{\begingroup\@sanitize@url \@url }%
\providecommand \@url [1]{\endgroup\@href {#1}{\urlprefix }}%
\providecommand \urlprefix  [0]{URL }%
\providecommand \Eprint [0]{\href }%
\providecommand \doibase [0]{https://doi.org/}%
\providecommand \selectlanguage [0]{\@gobble}%
\providecommand \bibinfo  [0]{\@secondoftwo}%
\providecommand \bibfield  [0]{\@secondoftwo}%
\providecommand \translation [1]{[#1]}%
\providecommand \BibitemOpen [0]{}%
\providecommand \bibitemStop [0]{}%
\providecommand \bibitemNoStop [0]{.\EOS\space}%
\providecommand \EOS [0]{\spacefactor3000\relax}%
\providecommand \BibitemShut  [1]{\csname bibitem#1\endcsname}%
\let\auto@bib@innerbib\@empty
\bibitem [{\citenamefont {Defenu}\ \emph
  {et~al.}(2019{\natexlab{a}})\citenamefont {Defenu}, \citenamefont {Enss},\
  and\ \citenamefont {Halimeh}}]{Defenu2019dynamical}%
  \BibitemOpen
  \bibfield  {author} {\bibinfo {author} {\bibfnamefont {N.}~\bibnamefont
  {Defenu}}, \bibinfo {author} {\bibfnamefont {T.}~\bibnamefont {Enss}},\ and\
  \bibinfo {author} {\bibfnamefont {J.~C.}\ \bibnamefont {Halimeh}},\
  }\bibfield  {title} {\bibinfo {title} {Dynamical criticality and domain-wall
  coupling in long-range hamiltonians},\ }\href
  {https://doi.org/10.1103/PhysRevB.100.014434} {\bibfield  {journal} {\bibinfo
   {journal} {Phys. Rev. B}\ }\textbf {\bibinfo {volume} {100}},\ \bibinfo
  {pages} {014434} (\bibinfo {year} {2019}{\natexlab{a}})}\BibitemShut
  {NoStop}%
\bibitem [{\citenamefont {Halimeh}\ \emph {et~al.}(2020)\citenamefont
  {Halimeh}, \citenamefont {Van~Damme}, \citenamefont {Zauner-Stauber},\ and\
  \citenamefont {Vanderstraeten}}]{Halimeh2020quasiparticle}%
  \BibitemOpen
  \bibfield  {author} {\bibinfo {author} {\bibfnamefont {J.~C.}\ \bibnamefont
  {Halimeh}}, \bibinfo {author} {\bibfnamefont {M.}~\bibnamefont {Van~Damme}},
  \bibinfo {author} {\bibfnamefont {V.}~\bibnamefont {Zauner-Stauber}},\ and\
  \bibinfo {author} {\bibfnamefont {L.}~\bibnamefont {Vanderstraeten}},\
  }\bibfield  {title} {\bibinfo {title} {Quasiparticle origin of dynamical
  quantum phase transitions},\ }\href
  {https://doi.org/10.1103/PhysRevResearch.2.033111} {\bibfield  {journal}
  {\bibinfo  {journal} {Phys. Rev. Research}\ }\textbf {\bibinfo {volume}
  {2}},\ \bibinfo {pages} {033111} (\bibinfo {year} {2020})}\BibitemShut
  {NoStop}%
\bibitem [{\citenamefont {Acevedo}\ \emph {et~al.}(2014)\citenamefont
  {Acevedo}, \citenamefont {Quiroga}, \citenamefont {Rodr\'{\i}guez},\ and\
  \citenamefont {Johnson}}]{Acevedo2014newdynamical}%
  \BibitemOpen
  \bibfield  {author} {\bibinfo {author} {\bibfnamefont {O.~L.}\ \bibnamefont
  {Acevedo}}, \bibinfo {author} {\bibfnamefont {L.}~\bibnamefont {Quiroga}},
  \bibinfo {author} {\bibfnamefont {F.~J.}\ \bibnamefont {Rodr\'{\i}guez}},\
  and\ \bibinfo {author} {\bibfnamefont {N.~F.}\ \bibnamefont {Johnson}},\
  }\bibfield  {title} {\bibinfo {title} {New dynamical scaling universality for
  quantum networks across adiabatic quantum phase transitions},\ }\href
  {https://doi.org/10.1103/PhysRevLett.112.030403} {\bibfield  {journal}
  {\bibinfo  {journal} {Phys. Rev. Lett.}\ }\textbf {\bibinfo {volume} {112}},\
  \bibinfo {pages} {030403} (\bibinfo {year} {2014})}\BibitemShut {NoStop}%
\bibitem [{\citenamefont {Hwang}\ \emph {et~al.}(2015)\citenamefont {Hwang},
  \citenamefont {Puebla},\ and\ \citenamefont {Plenio}}]{Hwang2015quantum}%
  \BibitemOpen
  \bibfield  {author} {\bibinfo {author} {\bibfnamefont {M.-J.}\ \bibnamefont
  {Hwang}}, \bibinfo {author} {\bibfnamefont {R.}~\bibnamefont {Puebla}},\ and\
  \bibinfo {author} {\bibfnamefont {M.~B.}\ \bibnamefont {Plenio}},\ }\bibfield
   {title} {\bibinfo {title} {Quantum phase transition and universal dynamics
  in the rabi model},\ }\href {https://doi.org/10.1103/PhysRevLett.115.180404}
  {\bibfield  {journal} {\bibinfo  {journal} {Phys. Rev. Lett.}\ }\textbf
  {\bibinfo {volume} {115}},\ \bibinfo {pages} {180404} (\bibinfo {year}
  {2015})}\BibitemShut {NoStop}%
\bibitem [{\citenamefont {Defenu}\ \emph {et~al.}(2018)\citenamefont {Defenu},
  \citenamefont {Enss}, \citenamefont {Kastner},\ and\ \citenamefont
  {Morigi}}]{Defenu2018dynamical}%
  \BibitemOpen
  \bibfield  {author} {\bibinfo {author} {\bibfnamefont {N.}~\bibnamefont
  {Defenu}}, \bibinfo {author} {\bibfnamefont {T.}~\bibnamefont {Enss}},
  \bibinfo {author} {\bibfnamefont {M.}~\bibnamefont {Kastner}},\ and\ \bibinfo
  {author} {\bibfnamefont {G.}~\bibnamefont {Morigi}},\ }\bibfield  {title}
  {\bibinfo {title} {Dynamical critical scaling of long-range interacting
  quantum magnets},\ }\href {https://doi.org/10.1103/PhysRevLett.121.240403}
  {\bibfield  {journal} {\bibinfo  {journal} {Phys. Rev. Lett.}\ }\textbf
  {\bibinfo {volume} {121}},\ \bibinfo {pages} {240403} (\bibinfo {year}
  {2018})}\BibitemShut {NoStop}%
\bibitem [{\citenamefont {Defenu}\ \emph
  {et~al.}(2019{\natexlab{b}})\citenamefont {Defenu}, \citenamefont {Morigi},
  \citenamefont {Dell'Anna},\ and\ \citenamefont {Enss}}]{defenu2019universal}%
  \BibitemOpen
  \bibfield  {author} {\bibinfo {author} {\bibfnamefont {N.}~\bibnamefont
  {Defenu}}, \bibinfo {author} {\bibfnamefont {G.}~\bibnamefont {Morigi}},
  \bibinfo {author} {\bibfnamefont {L.}~\bibnamefont {Dell'Anna}},\ and\
  \bibinfo {author} {\bibfnamefont {T.}~\bibnamefont {Enss}},\ }\bibfield
  {title} {\bibinfo {title} {Universal dynamical scaling of long-range
  topological superconductors},\ }\href
  {https://doi.org/10.1103/PhysRevB.100.184306} {\bibfield  {journal} {\bibinfo
   {journal} {Phys. Rev. B}\ }\textbf {\bibinfo {volume} {100}},\ \bibinfo
  {pages} {184306} (\bibinfo {year} {2019}{\natexlab{b}})}\BibitemShut
  {NoStop}%
\bibitem [{\citenamefont {Van~Regemortel}\ \emph {et~al.}(2016)\citenamefont
  {Van~Regemortel}, \citenamefont {Sels},\ and\ \citenamefont
  {Wouters}}]{Regemortel2016information}%
  \BibitemOpen
  \bibfield  {author} {\bibinfo {author} {\bibfnamefont {M.}~\bibnamefont
  {Van~Regemortel}}, \bibinfo {author} {\bibfnamefont {D.}~\bibnamefont
  {Sels}},\ and\ \bibinfo {author} {\bibfnamefont {M.}~\bibnamefont
  {Wouters}},\ }\bibfield  {title} {\bibinfo {title} {Information propagation
  and equilibration in long-range kitaev chains},\ }\href
  {https://doi.org/10.1103/PhysRevA.93.032311} {\bibfield  {journal} {\bibinfo
  {journal} {Phys. Rev. A}\ }\textbf {\bibinfo {volume} {93}},\ \bibinfo
  {pages} {032311} (\bibinfo {year} {2016})}\BibitemShut {NoStop}%
\bibitem [{\citenamefont {Tran}\ \emph {et~al.}(2020)\citenamefont {Tran},
  \citenamefont {Chen}, \citenamefont {Ehrenberg}, \citenamefont {Guo},
  \citenamefont {Deshpande}, \citenamefont {Hong}, \citenamefont {Gong},
  \citenamefont {Gorshkov},\ and\ \citenamefont {Lucas}}]{Tran2020hierarchy}%
  \BibitemOpen
  \bibfield  {author} {\bibinfo {author} {\bibfnamefont {M.~C.}\ \bibnamefont
  {Tran}}, \bibinfo {author} {\bibfnamefont {C.-F.}\ \bibnamefont {Chen}},
  \bibinfo {author} {\bibfnamefont {A.}~\bibnamefont {Ehrenberg}}, \bibinfo
  {author} {\bibfnamefont {A.~Y.}\ \bibnamefont {Guo}}, \bibinfo {author}
  {\bibfnamefont {A.}~\bibnamefont {Deshpande}}, \bibinfo {author}
  {\bibfnamefont {Y.}~\bibnamefont {Hong}}, \bibinfo {author} {\bibfnamefont
  {Z.-X.}\ \bibnamefont {Gong}}, \bibinfo {author} {\bibfnamefont {A.~V.}\
  \bibnamefont {Gorshkov}},\ and\ \bibinfo {author} {\bibfnamefont
  {A.}~\bibnamefont {Lucas}},\ }\bibfield  {title} {\bibinfo {title} {Hierarchy
  of linear light cones with long-range interactions},\ }\href
  {https://doi.org/10.1103/PhysRevX.10.031009} {\bibfield  {journal} {\bibinfo
  {journal} {Phys. Rev. X}\ }\textbf {\bibinfo {volume} {10}},\ \bibinfo
  {pages} {031009} (\bibinfo {year} {2020})}\BibitemShut {NoStop}%
\bibitem [{\citenamefont {Chen}\ and\ \citenamefont
  {Lucas}(2019)}]{Chen2019finite}%
  \BibitemOpen
  \bibfield  {author} {\bibinfo {author} {\bibfnamefont {C.-F.}\ \bibnamefont
  {Chen}}\ and\ \bibinfo {author} {\bibfnamefont {A.}~\bibnamefont {Lucas}},\
  }\bibfield  {title} {\bibinfo {title} {Finite speed of quantum scrambling
  with long range interactions},\ }\href
  {https://doi.org/10.1103/PhysRevLett.123.250605} {\bibfield  {journal}
  {\bibinfo  {journal} {Phys. Rev. Lett.}\ }\textbf {\bibinfo {volume} {123}},\
  \bibinfo {pages} {250605} (\bibinfo {year} {2019})}\BibitemShut {NoStop}%
\bibitem [{\citenamefont {Kuwahara}\ and\ \citenamefont
  {Saito}(2020)}]{Kuwahara2020strictly}%
  \BibitemOpen
  \bibfield  {author} {\bibinfo {author} {\bibfnamefont {T.}~\bibnamefont
  {Kuwahara}}\ and\ \bibinfo {author} {\bibfnamefont {K.}~\bibnamefont
  {Saito}},\ }\bibfield  {title} {\bibinfo {title} {Strictly linear light cones
  in long-range interacting systems of arbitrary dimensions},\ }\href
  {https://doi.org/10.1103/PhysRevX.10.031010} {\bibfield  {journal} {\bibinfo
  {journal} {Phys. Rev. X}\ }\textbf {\bibinfo {volume} {10}},\ \bibinfo
  {pages} {031010} (\bibinfo {year} {2020})}\BibitemShut {NoStop}%
\bibitem [{\citenamefont {Defenu}(2021)}]{Defenu2021metastability}%
  \BibitemOpen
  \bibfield  {author} {\bibinfo {author} {\bibfnamefont {N.}~\bibnamefont
  {Defenu}},\ }\bibfield  {title} {\bibinfo {title} {Metastability and discrete
  spectrum of long-range systems},\ }\href
  {https://doi.org/10.1073/pnas.2101785118} {\bibfield  {journal} {\bibinfo
  {journal} {Proceedings of the National Academy of Sciences}\ }\textbf
  {\bibinfo {volume} {118}},\ \bibinfo {pages} {e2101785118} (\bibinfo {year}
  {2021})},\ \Eprint
  {https://arxiv.org/abs/https://www.pnas.org/doi/pdf/10.1073/pnas.2101785118}
  {https://www.pnas.org/doi/pdf/10.1073/pnas.2101785118} \BibitemShut {NoStop}%
\bibitem [{\citenamefont {Giachetti}\ and\ \citenamefont
  {Defenu}(2023)}]{Giachetti2023entanglement}%
  \BibitemOpen
  \bibfield  {author} {\bibinfo {author} {\bibfnamefont {G.}~\bibnamefont
  {Giachetti}}\ and\ \bibinfo {author} {\bibfnamefont {N.}~\bibnamefont
  {Defenu}},\ }\bibfield  {title} {\bibinfo {title} {Entanglement propagation
  and dynamics in non-additive quantum systems},\ }\href
  {https://doi.org/10.1038/s41598-023-37984-3} {\bibfield  {journal} {\bibinfo
  {journal} {Scientific Reports}\ }\textbf {\bibinfo {volume} {13}},\ \bibinfo
  {pages} {12388} (\bibinfo {year} {2023})}\BibitemShut {NoStop}%
\bibitem [{\citenamefont {Xu}(2022)}]{xuPhysics2022}%
  \BibitemOpen
  \bibfield  {author} {\bibinfo {author} {\bibfnamefont {S.}~\bibnamefont
  {Xu}},\ }\bibfield  {title} {\bibinfo {title} {Long-range coupling affects
  entanglement dynamics},\ }\href {https://physics.aps.org/articles/v15/2}
  {\bibfield  {journal} {\bibinfo  {journal} {Physics}\ }\textbf {\bibinfo
  {volume} {15}},\ \bibinfo {pages} {2} (\bibinfo {year} {2022})}\BibitemShut
  {NoStop}%
\bibitem [{\citenamefont {Russomanno}\ \emph {et~al.}(2017)\citenamefont
  {Russomanno}, \citenamefont {Iemini}, \citenamefont {Dalmonte},\ and\
  \citenamefont {Fazio}}]{Russomanno2017Floquet}%
  \BibitemOpen
  \bibfield  {author} {\bibinfo {author} {\bibfnamefont {A.}~\bibnamefont
  {Russomanno}}, \bibinfo {author} {\bibfnamefont {F.}~\bibnamefont {Iemini}},
  \bibinfo {author} {\bibfnamefont {M.}~\bibnamefont {Dalmonte}},\ and\
  \bibinfo {author} {\bibfnamefont {R.}~\bibnamefont {Fazio}},\ }\bibfield
  {title} {\bibinfo {title} {Floquet time crystal in the lipkin-meshkov-glick
  model},\ }\href {https://doi.org/10.1103/PhysRevB.95.214307} {\bibfield
  {journal} {\bibinfo  {journal} {Phys. Rev. B}\ }\textbf {\bibinfo {volume}
  {95}},\ \bibinfo {pages} {214307} (\bibinfo {year} {2017})}\BibitemShut
  {NoStop}%
\bibitem [{\citenamefont {Pizzi}\ \emph {et~al.}(2021)\citenamefont {Pizzi},
  \citenamefont {Knolle},\ and\ \citenamefont
  {Nunnenkamp}}]{Pizzi2021higherorder}%
  \BibitemOpen
  \bibfield  {author} {\bibinfo {author} {\bibfnamefont {A.}~\bibnamefont
  {Pizzi}}, \bibinfo {author} {\bibfnamefont {J.}~\bibnamefont {Knolle}},\ and\
  \bibinfo {author} {\bibfnamefont {A.}~\bibnamefont {Nunnenkamp}},\ }\bibfield
   {title} {\bibinfo {title} {Higher-order and fractional discrete time
  crystals in clean long-range interacting systems},\ }\href
  {https://doi.org/10.1038/s41467-021-22583-5} {\bibfield  {journal} {\bibinfo
  {journal} {Nature Communications}\ }\textbf {\bibinfo {volume} {12}},\
  \bibinfo {pages} {2341} (\bibinfo {year} {2021})}\BibitemShut {NoStop}%
\bibitem [{\citenamefont {Giachetti}\ \emph {et~al.}(2023)\citenamefont
  {Giachetti}, \citenamefont {Solfanelli}, \citenamefont {Correale},\ and\
  \citenamefont {Defenu}}]{Giachetti2023fractal}%
  \BibitemOpen
  \bibfield  {author} {\bibinfo {author} {\bibfnamefont {G.}~\bibnamefont
  {Giachetti}}, \bibinfo {author} {\bibfnamefont {A.}~\bibnamefont
  {Solfanelli}}, \bibinfo {author} {\bibfnamefont {L.}~\bibnamefont
  {Correale}},\ and\ \bibinfo {author} {\bibfnamefont {N.}~\bibnamefont
  {Defenu}},\ }\bibfield  {title} {\bibinfo {title} {Fractal nature of
  high-order time crystal phases},\ }\href
  {https://doi.org/10.1103/PhysRevB.108.L140102} {\bibfield  {journal}
  {\bibinfo  {journal} {Phys. Rev. B}\ }\textbf {\bibinfo {volume} {108}},\
  \bibinfo {pages} {L140102} (\bibinfo {year} {2023})}\BibitemShut {NoStop}%
\bibitem [{\citenamefont {Solfanelli}\ \emph {et~al.}(2024)\citenamefont
  {Solfanelli}, \citenamefont {Ruffo}, \citenamefont {Succi},\ and\
  \citenamefont {Defenu}}]{Solfanelli2024stabilization}%
  \BibitemOpen
  \bibfield  {author} {\bibinfo {author} {\bibfnamefont {A.}~\bibnamefont
  {Solfanelli}}, \bibinfo {author} {\bibfnamefont {S.}~\bibnamefont {Ruffo}},
  \bibinfo {author} {\bibfnamefont {S.}~\bibnamefont {Succi}},\ and\ \bibinfo
  {author} {\bibfnamefont {N.}~\bibnamefont {Defenu}},\ }\bibfield  {title}
  {\bibinfo {title} {Stabilization of discrete time-crystalline response on a
  superconducting quantum computer by increasing the interaction range},\
  }\href {https://doi.org/10.1103/PhysRevResearch.6.013311} {\bibfield
  {journal} {\bibinfo  {journal} {Phys. Rev. Res.}\ }\textbf {\bibinfo {volume}
  {6}},\ \bibinfo {pages} {013311} (\bibinfo {year} {2024})}\BibitemShut
  {NoStop}%
\bibitem [{\citenamefont {Solfanelli}\ \emph
  {et~al.}(2023{\natexlab{a}})\citenamefont {Solfanelli}, \citenamefont
  {Ruffo}, \citenamefont {Succi},\ and\ \citenamefont
  {Defenu}}]{Solfanelli2023logarithmic}%
  \BibitemOpen
  \bibfield  {author} {\bibinfo {author} {\bibfnamefont {A.}~\bibnamefont
  {Solfanelli}}, \bibinfo {author} {\bibfnamefont {S.}~\bibnamefont {Ruffo}},
  \bibinfo {author} {\bibfnamefont {S.}~\bibnamefont {Succi}},\ and\ \bibinfo
  {author} {\bibfnamefont {N.}~\bibnamefont {Defenu}},\ }\bibfield  {title}
  {\bibinfo {title} {Logarithmic, fractal and volume-law entanglement in a
  kitaev chain with long-range hopping and pairing},\ }\href
  {https://doi.org/10.1007/JHEP05(2023)066} {\bibfield  {journal} {\bibinfo
  {journal} {Journal of High Energy Physics}\ }\textbf {\bibinfo {volume}
  {2023}},\ \bibinfo {pages} {66} (\bibinfo {year}
  {2023}{\natexlab{a}})}\BibitemShut {NoStop}%
\bibitem [{\citenamefont {Endres}\ \emph {et~al.}(2016)\citenamefont {Endres},
  \citenamefont {Bernien}, \citenamefont {Keesling}, \citenamefont {Levine},
  \citenamefont {Anschuetz}, \citenamefont {Krajenbrink}, \citenamefont
  {Senko}, \citenamefont {Vuletic}, \citenamefont {Greiner},\ and\
  \citenamefont {Lukin}}]{Endres2016atom}%
  \BibitemOpen
  \bibfield  {author} {\bibinfo {author} {\bibfnamefont {M.}~\bibnamefont
  {Endres}}, \bibinfo {author} {\bibfnamefont {H.}~\bibnamefont {Bernien}},
  \bibinfo {author} {\bibfnamefont {A.}~\bibnamefont {Keesling}}, \bibinfo
  {author} {\bibfnamefont {H.}~\bibnamefont {Levine}}, \bibinfo {author}
  {\bibfnamefont {E.~R.}\ \bibnamefont {Anschuetz}}, \bibinfo {author}
  {\bibfnamefont {A.}~\bibnamefont {Krajenbrink}}, \bibinfo {author}
  {\bibfnamefont {C.}~\bibnamefont {Senko}}, \bibinfo {author} {\bibfnamefont
  {V.}~\bibnamefont {Vuletic}}, \bibinfo {author} {\bibfnamefont
  {M.}~\bibnamefont {Greiner}},\ and\ \bibinfo {author} {\bibfnamefont {M.~D.}\
  \bibnamefont {Lukin}},\ }\bibfield  {title} {\bibinfo {title} {Atom-by-atom
  assembly of defect-free one-dimensional cold atom arrays},\ }\href
  {https://doi.org/10.1126/science.aah3752} {\bibfield  {journal} {\bibinfo
  {journal} {Science}\ }\textbf {\bibinfo {volume} {354}},\ \bibinfo {pages}
  {1024} (\bibinfo {year} {2016})},\ \Eprint
  {https://arxiv.org/abs/https://www.science.org/doi/pdf/10.1126/science.aah3752}
  {https://www.science.org/doi/pdf/10.1126/science.aah3752} \BibitemShut
  {NoStop}%
\bibitem [{\citenamefont {Labuhn}\ \emph {et~al.}(2016)\citenamefont {Labuhn},
  \citenamefont {Barredo}, \citenamefont {Ravets}, \citenamefont
  {de~L{\'e}s{\'e}leuc}, \citenamefont {Macr{\`i}}, \citenamefont {Lahaye},\
  and\ \citenamefont {Browaeys}}]{Labuhn2016tunable}%
  \BibitemOpen
  \bibfield  {author} {\bibinfo {author} {\bibfnamefont {H.}~\bibnamefont
  {Labuhn}}, \bibinfo {author} {\bibfnamefont {D.}~\bibnamefont {Barredo}},
  \bibinfo {author} {\bibfnamefont {S.}~\bibnamefont {Ravets}}, \bibinfo
  {author} {\bibfnamefont {S.}~\bibnamefont {de~L{\'e}s{\'e}leuc}}, \bibinfo
  {author} {\bibfnamefont {T.}~\bibnamefont {Macr{\`i}}}, \bibinfo {author}
  {\bibfnamefont {T.}~\bibnamefont {Lahaye}},\ and\ \bibinfo {author}
  {\bibfnamefont {A.}~\bibnamefont {Browaeys}},\ }\bibfield  {title} {\bibinfo
  {title} {Tunable two-dimensional arrays of single rydberg atoms for realizing
  quantum ising models},\ }\href {https://doi.org/10.1038/nature18274}
  {\bibfield  {journal} {\bibinfo  {journal} {Nature}\ }\textbf {\bibinfo
  {volume} {534}},\ \bibinfo {pages} {667} (\bibinfo {year}
  {2016})}\BibitemShut {NoStop}%
\bibitem [{\citenamefont {Zeiher}\ \emph {et~al.}(2017)\citenamefont {Zeiher},
  \citenamefont {Choi}, \citenamefont {Rubio-Abadal}, \citenamefont {Pohl},
  \citenamefont {van Bijnen}, \citenamefont {Bloch},\ and\ \citenamefont
  {Gross}}]{Zeiher2017coherent}%
  \BibitemOpen
  \bibfield  {author} {\bibinfo {author} {\bibfnamefont {J.}~\bibnamefont
  {Zeiher}}, \bibinfo {author} {\bibfnamefont {J.-y.}\ \bibnamefont {Choi}},
  \bibinfo {author} {\bibfnamefont {A.}~\bibnamefont {Rubio-Abadal}}, \bibinfo
  {author} {\bibfnamefont {T.}~\bibnamefont {Pohl}}, \bibinfo {author}
  {\bibfnamefont {R.}~\bibnamefont {van Bijnen}}, \bibinfo {author}
  {\bibfnamefont {I.}~\bibnamefont {Bloch}},\ and\ \bibinfo {author}
  {\bibfnamefont {C.}~\bibnamefont {Gross}},\ }\bibfield  {title} {\bibinfo
  {title} {Coherent many-body spin dynamics in a long-range interacting ising
  chain},\ }\href {https://doi.org/10.1103/PhysRevX.7.041063} {\bibfield
  {journal} {\bibinfo  {journal} {Phys. Rev. X}\ }\textbf {\bibinfo {volume}
  {7}},\ \bibinfo {pages} {041063} (\bibinfo {year} {2017})}\BibitemShut
  {NoStop}%
\bibitem [{\citenamefont {Hollerith}\ \emph {et~al.}(2022)\citenamefont
  {Hollerith}, \citenamefont {Srakaew}, \citenamefont {Wei}, \citenamefont
  {Rubio-Abadal}, \citenamefont {Adler}, \citenamefont {Weckesser},
  \citenamefont {Kruckenhauser}, \citenamefont {Walther}, \citenamefont {van
  Bijnen}, \citenamefont {Rui}, \citenamefont {Gross}, \citenamefont {Bloch},\
  and\ \citenamefont {Zeiher}}]{hollerith2022realizing}%
  \BibitemOpen
  \bibfield  {author} {\bibinfo {author} {\bibfnamefont {S.}~\bibnamefont
  {Hollerith}}, \bibinfo {author} {\bibfnamefont {K.}~\bibnamefont {Srakaew}},
  \bibinfo {author} {\bibfnamefont {D.}~\bibnamefont {Wei}}, \bibinfo {author}
  {\bibfnamefont {A.}~\bibnamefont {Rubio-Abadal}}, \bibinfo {author}
  {\bibfnamefont {D.}~\bibnamefont {Adler}}, \bibinfo {author} {\bibfnamefont
  {P.}~\bibnamefont {Weckesser}}, \bibinfo {author} {\bibfnamefont
  {A.}~\bibnamefont {Kruckenhauser}}, \bibinfo {author} {\bibfnamefont
  {V.}~\bibnamefont {Walther}}, \bibinfo {author} {\bibfnamefont
  {R.}~\bibnamefont {van Bijnen}}, \bibinfo {author} {\bibfnamefont
  {J.}~\bibnamefont {Rui}}, \bibinfo {author} {\bibfnamefont {C.}~\bibnamefont
  {Gross}}, \bibinfo {author} {\bibfnamefont {I.}~\bibnamefont {Bloch}},\ and\
  \bibinfo {author} {\bibfnamefont {J.}~\bibnamefont {Zeiher}},\ }\bibfield
  {title} {\bibinfo {title} {Realizing distance-selective interactions in a
  rydberg-dressed atom array},\ }\href
  {https://doi.org/10.1103/PhysRevLett.128.113602} {\bibfield  {journal}
  {\bibinfo  {journal} {Phys. Rev. Lett.}\ }\textbf {\bibinfo {volume} {128}},\
  \bibinfo {pages} {113602} (\bibinfo {year} {2022})}\BibitemShut {NoStop}%
\bibitem [{\citenamefont {Ritsch}\ \emph {et~al.}(2013)\citenamefont {Ritsch},
  \citenamefont {Domokos}, \citenamefont {Brennecke},\ and\ \citenamefont
  {Esslinger}}]{Ritsch2013cold}%
  \BibitemOpen
  \bibfield  {author} {\bibinfo {author} {\bibfnamefont {H.}~\bibnamefont
  {Ritsch}}, \bibinfo {author} {\bibfnamefont {P.}~\bibnamefont {Domokos}},
  \bibinfo {author} {\bibfnamefont {F.}~\bibnamefont {Brennecke}},\ and\
  \bibinfo {author} {\bibfnamefont {T.}~\bibnamefont {Esslinger}},\ }\bibfield
  {title} {\bibinfo {title} {Cold atoms in cavity-generated dynamical optical
  potentials},\ }\href {https://doi.org/10.1103/RevModPhys.85.553} {\bibfield
  {journal} {\bibinfo  {journal} {Rev. Mod. Phys.}\ }\textbf {\bibinfo {volume}
  {85}},\ \bibinfo {pages} {553} (\bibinfo {year} {2013})}\BibitemShut
  {NoStop}%
\bibitem [{\citenamefont {Farokh~Mivehvar}\ and\ \citenamefont
  {Ritsch}(2021)}]{Mivehvar2021cavity}%
  \BibitemOpen
  \bibfield  {author} {\bibinfo {author} {\bibfnamefont {T.~D.}\ \bibnamefont
  {Farokh~Mivehvar}, \bibfnamefont {Francesco~Piazza}}\ and\ \bibinfo {author}
  {\bibfnamefont {H.}~\bibnamefont {Ritsch}},\ }\bibfield  {title} {\bibinfo
  {title} {Cavity qed with quantum gases: new paradigms in many-body physics},\
  }\href {https://doi.org/10.1080/00018732.2021.1969727} {\bibfield  {journal}
  {\bibinfo  {journal} {Advances in Physics}\ }\textbf {\bibinfo {volume}
  {70}},\ \bibinfo {pages} {1} (\bibinfo {year} {2021})},\ \Eprint
  {https://arxiv.org/abs/https://doi.org/10.1080/00018732.2021.1969727}
  {https://doi.org/10.1080/00018732.2021.1969727} \BibitemShut {NoStop}%
\bibitem [{\citenamefont {Blatt}\ and\ \citenamefont
  {Roos}(2012)}]{Blatt2012quantum}%
  \BibitemOpen
  \bibfield  {author} {\bibinfo {author} {\bibfnamefont {R.}~\bibnamefont
  {Blatt}}\ and\ \bibinfo {author} {\bibfnamefont {C.~F.}\ \bibnamefont
  {Roos}},\ }\bibfield  {title} {\bibinfo {title} {Quantum simulations with
  trapped ions},\ }\href {https://doi.org/10.1038/nphys2252} {\bibfield
  {journal} {\bibinfo  {journal} {Nature Physics}\ }\textbf {\bibinfo {volume}
  {8}},\ \bibinfo {pages} {277} (\bibinfo {year} {2012})}\BibitemShut {NoStop}%
\bibitem [{\citenamefont {Monroe}\ \emph {et~al.}(2021)\citenamefont {Monroe},
  \citenamefont {Campbell}, \citenamefont {Duan}, \citenamefont {Gong},
  \citenamefont {Gorshkov}, \citenamefont {Hess}, \citenamefont {Islam},
  \citenamefont {Kim}, \citenamefont {Linke}, \citenamefont {Pagano},
  \citenamefont {Richerme}, \citenamefont {Senko},\ and\ \citenamefont
  {Yao}}]{Monroe2021programmable}%
  \BibitemOpen
  \bibfield  {author} {\bibinfo {author} {\bibfnamefont {C.}~\bibnamefont
  {Monroe}}, \bibinfo {author} {\bibfnamefont {W.~C.}\ \bibnamefont
  {Campbell}}, \bibinfo {author} {\bibfnamefont {L.-M.}\ \bibnamefont {Duan}},
  \bibinfo {author} {\bibfnamefont {Z.-X.}\ \bibnamefont {Gong}}, \bibinfo
  {author} {\bibfnamefont {A.~V.}\ \bibnamefont {Gorshkov}}, \bibinfo {author}
  {\bibfnamefont {P.~W.}\ \bibnamefont {Hess}}, \bibinfo {author}
  {\bibfnamefont {R.}~\bibnamefont {Islam}}, \bibinfo {author} {\bibfnamefont
  {K.}~\bibnamefont {Kim}}, \bibinfo {author} {\bibfnamefont {N.~M.}\
  \bibnamefont {Linke}}, \bibinfo {author} {\bibfnamefont {G.}~\bibnamefont
  {Pagano}}, \bibinfo {author} {\bibfnamefont {P.}~\bibnamefont {Richerme}},
  \bibinfo {author} {\bibfnamefont {C.}~\bibnamefont {Senko}},\ and\ \bibinfo
  {author} {\bibfnamefont {N.~Y.}\ \bibnamefont {Yao}},\ }\bibfield  {title}
  {\bibinfo {title} {Programmable quantum simulations of spin systems with
  trapped ions},\ }\href {https://doi.org/10.1103/RevModPhys.93.025001}
  {\bibfield  {journal} {\bibinfo  {journal} {Rev. Mod. Phys.}\ }\textbf
  {\bibinfo {volume} {93}},\ \bibinfo {pages} {025001} (\bibinfo {year}
  {2021})}\BibitemShut {NoStop}%
\bibitem [{\citenamefont {Defenu}\ \emph {et~al.}(2023)\citenamefont {Defenu},
  \citenamefont {Donner}, \citenamefont {Macr\`{\i}}, \citenamefont {Pagano},
  \citenamefont {Ruffo},\ and\ \citenamefont
  {Trombettoni}}]{Defenu2023Longrange}%
  \BibitemOpen
  \bibfield  {author} {\bibinfo {author} {\bibfnamefont {N.}~\bibnamefont
  {Defenu}}, \bibinfo {author} {\bibfnamefont {T.}~\bibnamefont {Donner}},
  \bibinfo {author} {\bibfnamefont {T.}~\bibnamefont {Macr\`{\i}}}, \bibinfo
  {author} {\bibfnamefont {G.}~\bibnamefont {Pagano}}, \bibinfo {author}
  {\bibfnamefont {S.}~\bibnamefont {Ruffo}},\ and\ \bibinfo {author}
  {\bibfnamefont {A.}~\bibnamefont {Trombettoni}},\ }\bibfield  {title}
  {\bibinfo {title} {Long-range interacting quantum systems},\ }\href
  {https://doi.org/10.1103/RevModPhys.95.035002} {\bibfield  {journal}
  {\bibinfo  {journal} {Rev. Mod. Phys.}\ }\textbf {\bibinfo {volume} {95}},\
  \bibinfo {pages} {035002} (\bibinfo {year} {2023})}\BibitemShut {NoStop}%
\bibitem [{\citenamefont {Smith}\ \emph {et~al.}(2022)\citenamefont {Smith},
  \citenamefont {Sinha},\ and\ \citenamefont {Jarzynski}}]{smith2022quantum}%
  \BibitemOpen
  \bibfield  {author} {\bibinfo {author} {\bibfnamefont {A.}~\bibnamefont
  {Smith}}, \bibinfo {author} {\bibfnamefont {K.}~\bibnamefont {Sinha}},\ and\
  \bibinfo {author} {\bibfnamefont {C.}~\bibnamefont {Jarzynski}},\ }\bibfield
  {title} {\bibinfo {title} {Quantum coherences and classical inhomogeneities
  as equivalent thermodynamics resources},\ }\bibfield  {journal} {\bibinfo
  {journal} {Entropy}\ }\textbf {\bibinfo {volume} {24}},\ \href
  {https://doi.org/10.3390/e24040474} {10.3390/e24040474} (\bibinfo {year}
  {2022})\BibitemShut {NoStop}%
\bibitem [{\citenamefont {Goold}\ \emph {et~al.}(2018)\citenamefont {Goold},
  \citenamefont {Plastina}, \citenamefont {Gambassi},\ and\ \citenamefont
  {Silva}}]{Goold2018therole}%
  \BibitemOpen
  \bibfield  {author} {\bibinfo {author} {\bibfnamefont {J.}~\bibnamefont
  {Goold}}, \bibinfo {author} {\bibfnamefont {F.}~\bibnamefont {Plastina}},
  \bibinfo {author} {\bibfnamefont {A.}~\bibnamefont {Gambassi}},\ and\
  \bibinfo {author} {\bibfnamefont {A.}~\bibnamefont {Silva}},\ }\bibinfo
  {title} {The role of quantum work statistics in many-body physics},\ in\
  \href {https://doi.org/10.1007/978-3-319-99046-0_13} {\emph {\bibinfo
  {booktitle} {Thermodynamics in the Quantum Regime: Fundamental Aspects and
  New Directions}}},\ \bibinfo {editor} {edited by\ \bibinfo {editor}
  {\bibfnamefont {F.}~\bibnamefont {Binder}}, \bibinfo {editor} {\bibfnamefont
  {L.~A.}\ \bibnamefont {Correa}}, \bibinfo {editor} {\bibfnamefont
  {C.}~\bibnamefont {Gogolin}}, \bibinfo {editor} {\bibfnamefont
  {J.}~\bibnamefont {Anders}},\ and\ \bibinfo {editor} {\bibfnamefont
  {G.}~\bibnamefont {Adesso}}}\ (\bibinfo  {publisher} {Springer International
  Publishing},\ \bibinfo {address} {Cham},\ \bibinfo {year} {2018})\ pp.\
  \bibinfo {pages} {317--336}\BibitemShut {NoStop}%
\bibitem [{\citenamefont {Heyl}(2018)}]{Heyl2018dynamical}%
  \BibitemOpen
  \bibfield  {author} {\bibinfo {author} {\bibfnamefont {M.}~\bibnamefont
  {Heyl}},\ }\bibfield  {title} {\bibinfo {title} {Dynamical quantum phase
  transitions: a review},\ }\href {https://doi.org/10.1088/1361-6633/aaaf9a}
  {\bibfield  {journal} {\bibinfo  {journal} {Reports on Progress in Physics}\
  }\textbf {\bibinfo {volume} {81}},\ \bibinfo {pages} {054001} (\bibinfo
  {year} {2018})}\BibitemShut {NoStop}%
\bibitem [{\citenamefont {Campisi}\ and\ \citenamefont
  {Goold}(2017)}]{campisi2017thermodynamics}%
  \BibitemOpen
  \bibfield  {author} {\bibinfo {author} {\bibfnamefont {M.}~\bibnamefont
  {Campisi}}\ and\ \bibinfo {author} {\bibfnamefont {J.}~\bibnamefont
  {Goold}},\ }\bibfield  {title} {\bibinfo {title} {Thermodynamics of quantum
  information scrambling},\ }\href {https://doi.org/10.1103/PhysRevE.95.062127}
  {\bibfield  {journal} {\bibinfo  {journal} {Phys. Rev. E}\ }\textbf {\bibinfo
  {volume} {95}},\ \bibinfo {pages} {062127} (\bibinfo {year}
  {2017})}\BibitemShut {NoStop}%
\bibitem [{\citenamefont {Deffner}\ and\ \citenamefont
  {Campbell}(2019)}]{Deffner19Book}%
  \BibitemOpen
  \bibfield  {author} {\bibinfo {author} {\bibfnamefont {S.}~\bibnamefont
  {Deffner}}\ and\ \bibinfo {author} {\bibfnamefont {S.}~\bibnamefont
  {Campbell}},\ }\bibfield  {title} {\bibinfo {title} {Thermodynamics of
  quantum systems},\ }in\ \href {https://doi.org/10.1088/2053-2571/ab21c6ch2}
  {\emph {\bibinfo {booktitle} {Quantum Thermodynamics}}},\ \bibinfo {series
  and number} {2053-2571}\ (\bibinfo  {publisher} {Morgan \& Claypool
  Publishers},\ \bibinfo {year} {2019})\ pp.\ \bibinfo {pages} {2--1 to
  2--37}\BibitemShut {NoStop}%
\bibitem [{\citenamefont {Fermi}(1956)}]{Fermi1956}%
  \BibitemOpen
  \bibfield  {author} {\bibinfo {author} {\bibfnamefont {E.}~\bibnamefont
  {Fermi}},\ }\href@noop {} {\emph {\bibinfo {title} {Thermodynamics}}}\
  (\bibinfo  {publisher} {Dover Publications},\ \bibinfo {year}
  {1956})\BibitemShut {NoStop}%
\bibitem [{\citenamefont {Solfanelli}\ \emph
  {et~al.}(2023{\natexlab{b}})\citenamefont {Solfanelli}, \citenamefont
  {Giachetti}, \citenamefont {Campisi}, \citenamefont {Ruffo},\ and\
  \citenamefont {Defenu}}]{Solfanelli2023quantum}%
  \BibitemOpen
  \bibfield  {author} {\bibinfo {author} {\bibfnamefont {A.}~\bibnamefont
  {Solfanelli}}, \bibinfo {author} {\bibfnamefont {G.}~\bibnamefont
  {Giachetti}}, \bibinfo {author} {\bibfnamefont {M.}~\bibnamefont {Campisi}},
  \bibinfo {author} {\bibfnamefont {S.}~\bibnamefont {Ruffo}},\ and\ \bibinfo
  {author} {\bibfnamefont {N.}~\bibnamefont {Defenu}},\ }\bibfield  {title}
  {\bibinfo {title} {Quantum heat engine with long-range advantages},\ }\href
  {https://doi.org/10.1088/1367-2630/acc04e} {\bibfield  {journal} {\bibinfo
  {journal} {New Journal of Physics}\ }\textbf {\bibinfo {volume} {25}},\
  \bibinfo {pages} {033030} (\bibinfo {year} {2023}{\natexlab{b}})}\BibitemShut
  {NoStop}%
\bibitem [{\citenamefont {Solfanelli}\ and\ \citenamefont
  {Defenu}(2024)}]{solfanelli2024universality}%
  \BibitemOpen
  \bibfield  {author} {\bibinfo {author} {\bibfnamefont {A.}~\bibnamefont
  {Solfanelli}}\ and\ \bibinfo {author} {\bibfnamefont {N.}~\bibnamefont
  {Defenu}},\ }\bibfield  {title} {\bibinfo {title} {Universality in long-range
  interacting systems: the effective dimension approach},\ }\href
  {https://arxiv.org/abs/2406.14651} {\bibfield  {journal} {\bibinfo  {journal}
  {arXiv}\ }\textbf {\bibinfo {volume} {2406}},\ \bibinfo {pages} {14651}
  (\bibinfo {year} {2024})}\BibitemShut {NoStop}%
\bibitem [{\citenamefont {Mermin}\ and\ \citenamefont
  {Wagner}(1966)}]{Mermin1966absence}%
  \BibitemOpen
  \bibfield  {author} {\bibinfo {author} {\bibfnamefont {N.~D.}\ \bibnamefont
  {Mermin}}\ and\ \bibinfo {author} {\bibfnamefont {H.}~\bibnamefont
  {Wagner}},\ }\bibfield  {title} {\bibinfo {title} {Absence of ferromagnetism
  or antiferromagnetism in one- or two-dimensional isotropic heisenberg
  models},\ }\href {https://doi.org/10.1103/PhysRevLett.17.1133} {\bibfield
  {journal} {\bibinfo  {journal} {Phys. Rev. Lett.}\ }\textbf {\bibinfo
  {volume} {17}},\ \bibinfo {pages} {1133} (\bibinfo {year}
  {1966})}\BibitemShut {NoStop}%
\bibitem [{\citenamefont {Halperin}(2018)}]{halperin2018hohenberg}%
  \BibitemOpen
  \bibfield  {author} {\bibinfo {author} {\bibfnamefont {B.~I.}\ \bibnamefont
  {Halperin}},\ }\bibfield  {title} {\bibinfo {title} {On the
  hohenberg–mermin–wagner theorem and its limitations},\ }\href
  {https://doi.org/10.1007/s10955-018-2202-y} {\bibfield  {journal} {\bibinfo
  {journal} {Journal of Statistical Physics}\ }\textbf {\bibinfo {volume}
  {175}},\ \bibinfo {pages} {521–529} (\bibinfo {year} {2018})}\BibitemShut
  {NoStop}%
\bibitem [{\citenamefont {Chen}\ \emph {et~al.}(2023)\citenamefont {Chen},
  \citenamefont {Bornet}, \citenamefont {Bintz}, \citenamefont {Emperauger},
  \citenamefont {Leclerc}, \citenamefont {Liu}, \citenamefont {Scholl},
  \citenamefont {Barredo}, \citenamefont {Hauschild}, \citenamefont
  {Chatterjee}, \citenamefont {Schuler}, \citenamefont {L{\"a}uchli},
  \citenamefont {Zaletel}, \citenamefont {Lahaye}, \citenamefont {Yao},\ and\
  \citenamefont {Browaeys}}]{chen2023continuous}%
  \BibitemOpen
  \bibfield  {author} {\bibinfo {author} {\bibfnamefont {C.}~\bibnamefont
  {Chen}}, \bibinfo {author} {\bibfnamefont {G.}~\bibnamefont {Bornet}},
  \bibinfo {author} {\bibfnamefont {M.}~\bibnamefont {Bintz}}, \bibinfo
  {author} {\bibfnamefont {G.}~\bibnamefont {Emperauger}}, \bibinfo {author}
  {\bibfnamefont {L.}~\bibnamefont {Leclerc}}, \bibinfo {author} {\bibfnamefont
  {V.~S.}\ \bibnamefont {Liu}}, \bibinfo {author} {\bibfnamefont
  {P.}~\bibnamefont {Scholl}}, \bibinfo {author} {\bibfnamefont
  {D.}~\bibnamefont {Barredo}}, \bibinfo {author} {\bibfnamefont
  {J.}~\bibnamefont {Hauschild}}, \bibinfo {author} {\bibfnamefont
  {S.}~\bibnamefont {Chatterjee}}, \bibinfo {author} {\bibfnamefont
  {M.}~\bibnamefont {Schuler}}, \bibinfo {author} {\bibfnamefont {A.~M.}\
  \bibnamefont {L{\"a}uchli}}, \bibinfo {author} {\bibfnamefont {M.~P.}\
  \bibnamefont {Zaletel}}, \bibinfo {author} {\bibfnamefont {T.}~\bibnamefont
  {Lahaye}}, \bibinfo {author} {\bibfnamefont {N.~Y.}\ \bibnamefont {Yao}},\
  and\ \bibinfo {author} {\bibfnamefont {A.}~\bibnamefont {Browaeys}},\
  }\bibfield  {title} {\bibinfo {title} {Continuous symmetry breaking in a
  two-dimensional rydberg array},\ }\href
  {https://doi.org/10.1038/s41586-023-05859-2} {\bibfield  {journal} {\bibinfo
  {journal} {Nature}\ }\textbf {\bibinfo {volume} {616}},\ \bibinfo {pages}
  {691} (\bibinfo {year} {2023})}\BibitemShut {NoStop}%
\bibitem [{\citenamefont {Feng}\ \emph {et~al.}(2023)\citenamefont {Feng},
  \citenamefont {Katz}, \citenamefont {Haack}, \citenamefont {Maghrebi},
  \citenamefont {Gorshkov}, \citenamefont {Gong}, \citenamefont {Cetina},\ and\
  \citenamefont {Monroe}}]{feng2023continuous}%
  \BibitemOpen
  \bibfield  {author} {\bibinfo {author} {\bibfnamefont {L.}~\bibnamefont
  {Feng}}, \bibinfo {author} {\bibfnamefont {O.}~\bibnamefont {Katz}}, \bibinfo
  {author} {\bibfnamefont {C.}~\bibnamefont {Haack}}, \bibinfo {author}
  {\bibfnamefont {M.}~\bibnamefont {Maghrebi}}, \bibinfo {author}
  {\bibfnamefont {A.~V.}\ \bibnamefont {Gorshkov}}, \bibinfo {author}
  {\bibfnamefont {Z.}~\bibnamefont {Gong}}, \bibinfo {author} {\bibfnamefont
  {M.}~\bibnamefont {Cetina}},\ and\ \bibinfo {author} {\bibfnamefont
  {C.}~\bibnamefont {Monroe}},\ }\bibfield  {title} {\bibinfo {title}
  {Continuous symmetry breaking in a trapped-ion spin chain},\ }\href
  {https://doi.org/10.1038/s41586-023-06656-7} {\bibfield  {journal} {\bibinfo
  {journal} {Nature}\ }\textbf {\bibinfo {volume} {623}},\ \bibinfo {pages}
  {713} (\bibinfo {year} {2023})}\BibitemShut {NoStop}%
\bibitem [{\citenamefont {Angelini}\ \emph {et~al.}(2014)\citenamefont
  {Angelini}, \citenamefont {Parisi},\ and\ \citenamefont
  {Ricci-Tersenghi}}]{Angelini2014relations}%
  \BibitemOpen
  \bibfield  {author} {\bibinfo {author} {\bibfnamefont {M.~C.}\ \bibnamefont
  {Angelini}}, \bibinfo {author} {\bibfnamefont {G.}~\bibnamefont {Parisi}},\
  and\ \bibinfo {author} {\bibfnamefont {F.}~\bibnamefont {Ricci-Tersenghi}},\
  }\bibfield  {title} {\bibinfo {title} {Relations between short-range and
  long-range ising models},\ }\href
  {https://doi.org/10.1103/PhysRevE.89.062120} {\bibfield  {journal} {\bibinfo
  {journal} {Phys. Rev. E}\ }\textbf {\bibinfo {volume} {89}},\ \bibinfo
  {pages} {062120} (\bibinfo {year} {2014})}\BibitemShut {NoStop}%
\bibitem [{SM()}]{SM}%
  \BibitemOpen
  \href@noop {} {}\bibinfo {howpublished} {See Supplemental
  Material.}\BibitemShut {Stop}%
\bibitem [{\citenamefont {Talkner}\ \emph {et~al.}(2007)\citenamefont
  {Talkner}, \citenamefont {Lutz},\ and\ \citenamefont
  {H\"anggi}}]{Talkner2007fluctuation}%
  \BibitemOpen
  \bibfield  {author} {\bibinfo {author} {\bibfnamefont {P.}~\bibnamefont
  {Talkner}}, \bibinfo {author} {\bibfnamefont {E.}~\bibnamefont {Lutz}},\ and\
  \bibinfo {author} {\bibfnamefont {P.}~\bibnamefont {H\"anggi}},\ }\bibfield
  {title} {\bibinfo {title} {Fluctuation theorems: Work is not an observable},\
  }\href {https://doi.org/10.1103/PhysRevE.75.050102} {\bibfield  {journal}
  {\bibinfo  {journal} {Phys. Rev. E}\ }\textbf {\bibinfo {volume} {75}},\
  \bibinfo {pages} {050102} (\bibinfo {year} {2007})}\BibitemShut {NoStop}%
\bibitem [{\citenamefont {Santini}\ \emph {et~al.}(2023)\citenamefont
  {Santini}, \citenamefont {Solfanelli}, \citenamefont {Gherardini},\ and\
  \citenamefont {Collura}}]{Santini2023work}%
  \BibitemOpen
  \bibfield  {author} {\bibinfo {author} {\bibfnamefont {A.}~\bibnamefont
  {Santini}}, \bibinfo {author} {\bibfnamefont {A.}~\bibnamefont {Solfanelli}},
  \bibinfo {author} {\bibfnamefont {S.}~\bibnamefont {Gherardini}},\ and\
  \bibinfo {author} {\bibfnamefont {M.}~\bibnamefont {Collura}},\ }\bibfield
  {title} {\bibinfo {title} {Work statistics, quantum signatures, and enhanced
  work extraction in quadratic fermionic models},\ }\href
  {https://doi.org/10.1103/PhysRevB.108.104308} {\bibfield  {journal} {\bibinfo
   {journal} {Phys. Rev. B}\ }\textbf {\bibinfo {volume} {108}},\ \bibinfo
  {pages} {104308} (\bibinfo {year} {2023})}\BibitemShut {NoStop}%
\bibitem [{\citenamefont {Gambassi}\ and\ \citenamefont
  {Silva}(2012)}]{Gambassi2012largedeviations}%
  \BibitemOpen
  \bibfield  {author} {\bibinfo {author} {\bibfnamefont {A.}~\bibnamefont
  {Gambassi}}\ and\ \bibinfo {author} {\bibfnamefont {A.}~\bibnamefont
  {Silva}},\ }\bibfield  {title} {\bibinfo {title} {Large deviations and
  universality in quantum quenches},\ }\href
  {https://doi.org/10.1103/PhysRevLett.109.250602} {\bibfield  {journal}
  {\bibinfo  {journal} {Phys. Rev. Lett.}\ }\textbf {\bibinfo {volume} {109}},\
  \bibinfo {pages} {250602} (\bibinfo {year} {2012})}\BibitemShut {NoStop}%
\bibitem [{\citenamefont {Gambassi}\ and\ \citenamefont
  {Silva}(2011)}]{gambassi2011statistics}%
  \BibitemOpen
  \bibfield  {author} {\bibinfo {author} {\bibfnamefont {A.}~\bibnamefont
  {Gambassi}}\ and\ \bibinfo {author} {\bibfnamefont {A.}~\bibnamefont
  {Silva}},\ }\href@noop {} {\bibinfo {title} {Statistics of the work in
  quantum quenches, universality and the critical casimir effect}} (\bibinfo
  {year} {2011}),\ \Eprint {https://arxiv.org/abs/1106.2671} {arXiv:1106.2671}
  \BibitemShut {NoStop}%
\bibitem [{\citenamefont {Krech}(1994)}]{krech1994thecasimir}%
  \BibitemOpen
  \bibfield  {author} {\bibinfo {author} {\bibfnamefont {M.}~\bibnamefont
  {Krech}},\ }\href {https://doi.org/10.1142/2434} {\emph {\bibinfo {title}
  {The Casimir Effect in Critical Systems}}}\ (\bibinfo  {publisher} {WORLD
  SCIENTIFIC},\ \bibinfo {year} {1994})\ \Eprint
  {https://arxiv.org/abs/https://www.worldscientific.com/doi/pdf/10.1142/2434}
  {https://www.worldscientific.com/doi/pdf/10.1142/2434} \BibitemShut {NoStop}%
\bibitem [{\citenamefont {Krech}(1999)}]{krech1999fluctuation}%
  \BibitemOpen
  \bibfield  {author} {\bibinfo {author} {\bibfnamefont {M.}~\bibnamefont
  {Krech}},\ }\bibfield  {title} {\bibinfo {title} {Fluctuation-induced forces
  in critical fluids},\ }\href {https://doi.org/10.1088/0953-8984/11/37/201}
  {\bibfield  {journal} {\bibinfo  {journal} {Journal of Physics: Condensed
  Matter}\ }\textbf {\bibinfo {volume} {11}},\ \bibinfo {pages} {R391}
  (\bibinfo {year} {1999})}\BibitemShut {NoStop}%
\bibitem [{\citenamefont {Gambassi}(2009)}]{gambassi2009thecasimir}%
  \BibitemOpen
  \bibfield  {author} {\bibinfo {author} {\bibfnamefont {A.}~\bibnamefont
  {Gambassi}},\ }\bibfield  {title} {\bibinfo {title} {The casimir effect: From
  quantum to critical fluctuations},\ }\href
  {https://doi.org/10.1088/1742-6596/161/1/012037} {\bibfield  {journal}
  {\bibinfo  {journal} {Journal of Physics: Conference Series}\ }\textbf
  {\bibinfo {volume} {161}},\ \bibinfo {pages} {012037} (\bibinfo {year}
  {2009})}\BibitemShut {NoStop}%
\bibitem [{\citenamefont {Binder}\ \emph {et~al.}(2002)\citenamefont {Binder},
  \citenamefont {Haas},\ and\ \citenamefont {Schmid}}]{binder2002critical}%
  \BibitemOpen
  \bibfield  {author} {\bibinfo {author} {\bibfnamefont {K.}~\bibnamefont
  {Binder}}, \bibinfo {author} {\bibfnamefont {F.~F.}\ \bibnamefont {Haas}},\
  and\ \bibinfo {author} {\bibfnamefont {F.}~\bibnamefont {Schmid}},\
  }\bibfield  {title} {\bibinfo {title} {Critical phenomena at the surface of
  systems undergoing a bulk first order transition: Are they understood?},\
  }in\ \href@noop {} {\emph {\bibinfo {booktitle} {Computer Simulation Studies
  in Condensed-Matter Physics XIV}}},\ \bibinfo {editor} {edited by\ \bibinfo
  {editor} {\bibfnamefont {D.~P.}\ \bibnamefont {Landau}}, \bibinfo {editor}
  {\bibfnamefont {S.~P.}\ \bibnamefont {Lewis}},\ and\ \bibinfo {editor}
  {\bibfnamefont {H.-B.}\ \bibnamefont {Sch{\"u}ttler}}}\ (\bibinfo
  {publisher} {Springer Berlin Heidelberg},\ \bibinfo {address} {Berlin,
  Heidelberg},\ \bibinfo {year} {2002})\ pp.\ \bibinfo {pages}
  {85--96}\BibitemShut {NoStop}%
\bibitem [{\citenamefont {Touchette}(2009)}]{touchette2009largedeviation}%
  \BibitemOpen
  \bibfield  {author} {\bibinfo {author} {\bibfnamefont {H.}~\bibnamefont
  {Touchette}},\ }\bibfield  {title} {\bibinfo {title} {The large deviation
  approach to statistical mechanics},\ }\href
  {https://doi.org/https://doi.org/10.1016/j.physrep.2009.05.002} {\bibfield
  {journal} {\bibinfo  {journal} {Physics Reports}\ }\textbf {\bibinfo {volume}
  {478}},\ \bibinfo {pages} {1} (\bibinfo {year} {2009})}\BibitemShut {NoStop}%
\bibitem [{\citenamefont {Silva}(2008)}]{silva2008statistics}%
  \BibitemOpen
  \bibfield  {author} {\bibinfo {author} {\bibfnamefont {A.}~\bibnamefont
  {Silva}},\ }\bibfield  {title} {\bibinfo {title} {Statistics of the work done
  on a quantum critical system by quenching a control parameter},\ }\href
  {https://doi.org/10.1103/PhysRevLett.101.120603} {\bibfield  {journal}
  {\bibinfo  {journal} {Phys. Rev. Lett.}\ }\textbf {\bibinfo {volume} {101}},\
  \bibinfo {pages} {120603} (\bibinfo {year} {2008})}\BibitemShut {NoStop}%
\bibitem [{\citenamefont {Zawadzki}\ \emph {et~al.}(2023)\citenamefont
  {Zawadzki}, \citenamefont {Kiely}, \citenamefont {Landi},\ and\ \citenamefont
  {Campbell}}]{zawadzki2023nongaussian}%
  \BibitemOpen
  \bibfield  {author} {\bibinfo {author} {\bibfnamefont {K.}~\bibnamefont
  {Zawadzki}}, \bibinfo {author} {\bibfnamefont {A.}~\bibnamefont {Kiely}},
  \bibinfo {author} {\bibfnamefont {G.~T.}\ \bibnamefont {Landi}},\ and\
  \bibinfo {author} {\bibfnamefont {S.}~\bibnamefont {Campbell}},\ }\bibfield
  {title} {\bibinfo {title} {Non-gaussian work statistics at finite-time
  driving},\ }\href {https://doi.org/10.1103/PhysRevA.107.012209} {\bibfield
  {journal} {\bibinfo  {journal} {Phys. Rev. A}\ }\textbf {\bibinfo {volume}
  {107}},\ \bibinfo {pages} {012209} (\bibinfo {year} {2023})}\BibitemShut
  {NoStop}%
\bibitem [{\citenamefont {Polkovnikov}(2005)}]{Polkovnikov2005universal}%
  \BibitemOpen
  \bibfield  {author} {\bibinfo {author} {\bibfnamefont {A.}~\bibnamefont
  {Polkovnikov}},\ }\bibfield  {title} {\bibinfo {title} {Universal adiabatic
  dynamics in the vicinity of a quantum critical point},\ }\href
  {https://doi.org/10.1103/PhysRevB.72.161201} {\bibfield  {journal} {\bibinfo
  {journal} {Phys. Rev. B}\ }\textbf {\bibinfo {volume} {72}},\ \bibinfo
  {pages} {161201} (\bibinfo {year} {2005})}\BibitemShut {NoStop}%
\bibitem [{\citenamefont {De~Grandi}\ and\ \citenamefont
  {Polkovnikov}(2010)}]{DeGrandi2010adiabatic}%
  \BibitemOpen
  \bibfield  {author} {\bibinfo {author} {\bibfnamefont {C.}~\bibnamefont
  {De~Grandi}}\ and\ \bibinfo {author} {\bibfnamefont {A.}~\bibnamefont
  {Polkovnikov}},\ }\bibinfo {title} {Adiabatic perturbation theory: From
  landau--zener problem to quenching through a quantum critical point},\ in\
  \href {https://doi.org/10.1007/978-3-642-11470-0_4} {\emph {\bibinfo
  {booktitle} {Quantum Quenching, Annealing and Computation}}},\ \bibinfo
  {editor} {edited by\ \bibinfo {editor} {\bibfnamefont {A.~K.}\ \bibnamefont
  {Chandra}}, \bibinfo {editor} {\bibfnamefont {A.}~\bibnamefont {Das}},\ and\
  \bibinfo {editor} {\bibfnamefont {B.~K.}\ \bibnamefont {Chakrabarti}}}\
  (\bibinfo  {publisher} {Springer Berlin Heidelberg},\ \bibinfo {address}
  {Berlin, Heidelberg},\ \bibinfo {year} {2010})\ pp.\ \bibinfo {pages}
  {75--114}\BibitemShut {NoStop}%
\bibitem [{\citenamefont {del Campo}(2018)}]{delcampo2018universal}%
  \BibitemOpen
  \bibfield  {author} {\bibinfo {author} {\bibfnamefont {A.}~\bibnamefont {del
  Campo}},\ }\bibfield  {title} {\bibinfo {title} {Universal statistics of
  topological defects formed in a quantum phase transition},\ }\href
  {https://doi.org/10.1103/PhysRevLett.121.200601} {\bibfield  {journal}
  {\bibinfo  {journal} {Phys. Rev. Lett.}\ }\textbf {\bibinfo {volume} {121}},\
  \bibinfo {pages} {200601} (\bibinfo {year} {2018})}\BibitemShut {NoStop}%
\bibitem [{\citenamefont {Fei}\ \emph {et~al.}(2020)\citenamefont {Fei},
  \citenamefont {Freitas}, \citenamefont {Cavina}, \citenamefont {Quan},\ and\
  \citenamefont {Esposito}}]{Fei2020work}%
  \BibitemOpen
  \bibfield  {author} {\bibinfo {author} {\bibfnamefont {Z.}~\bibnamefont
  {Fei}}, \bibinfo {author} {\bibfnamefont {N.}~\bibnamefont {Freitas}},
  \bibinfo {author} {\bibfnamefont {V.}~\bibnamefont {Cavina}}, \bibinfo
  {author} {\bibfnamefont {H.~T.}\ \bibnamefont {Quan}},\ and\ \bibinfo
  {author} {\bibfnamefont {M.}~\bibnamefont {Esposito}},\ }\bibfield  {title}
  {\bibinfo {title} {Work statistics across a quantum phase transition},\
  }\href {https://doi.org/10.1103/PhysRevLett.124.170603} {\bibfield  {journal}
  {\bibinfo  {journal} {Phys. Rev. Lett.}\ }\textbf {\bibinfo {volume} {124}},\
  \bibinfo {pages} {170603} (\bibinfo {year} {2020})}\BibitemShut {NoStop}%
\bibitem [{\citenamefont {Balducci}\ \emph {et~al.}(2023)\citenamefont
  {Balducci}, \citenamefont {Beau}, \citenamefont {Yang}, \citenamefont
  {Gambassi},\ and\ \citenamefont {del Campo}}]{balducci2023large}%
  \BibitemOpen
  \bibfield  {author} {\bibinfo {author} {\bibfnamefont {F.}~\bibnamefont
  {Balducci}}, \bibinfo {author} {\bibfnamefont {M.}~\bibnamefont {Beau}},
  \bibinfo {author} {\bibfnamefont {J.}~\bibnamefont {Yang}}, \bibinfo {author}
  {\bibfnamefont {A.}~\bibnamefont {Gambassi}},\ and\ \bibinfo {author}
  {\bibfnamefont {A.}~\bibnamefont {del Campo}},\ }\bibfield  {title} {\bibinfo
  {title} {Large deviations beyond the kibble-zurek mechanism},\ }\href
  {https://doi.org/10.1103/PhysRevLett.131.230401} {\bibfield  {journal}
  {\bibinfo  {journal} {Phys. Rev. Lett.}\ }\textbf {\bibinfo {volume} {131}},\
  \bibinfo {pages} {230401} (\bibinfo {year} {2023})}\BibitemShut {NoStop}%
\bibitem [{\citenamefont {Polkovnikov}\ \emph {et~al.}(2011)\citenamefont
  {Polkovnikov}, \citenamefont {Sengupta}, \citenamefont {Silva},\ and\
  \citenamefont {Vengalattore}}]{Polkovnikov2011colloquium}%
  \BibitemOpen
  \bibfield  {author} {\bibinfo {author} {\bibfnamefont {A.}~\bibnamefont
  {Polkovnikov}}, \bibinfo {author} {\bibfnamefont {K.}~\bibnamefont
  {Sengupta}}, \bibinfo {author} {\bibfnamefont {A.}~\bibnamefont {Silva}},\
  and\ \bibinfo {author} {\bibfnamefont {M.}~\bibnamefont {Vengalattore}},\
  }\bibfield  {title} {\bibinfo {title} {Colloquium: Nonequilibrium dynamics of
  closed interacting quantum systems},\ }\href
  {https://doi.org/10.1103/RevModPhys.83.863} {\bibfield  {journal} {\bibinfo
  {journal} {Rev. Mod. Phys.}\ }\textbf {\bibinfo {volume} {83}},\ \bibinfo
  {pages} {863} (\bibinfo {year} {2011})}\BibitemShut {NoStop}%
\bibitem [{\citenamefont {El-Showk}\ \emph {et~al.}(2014)\citenamefont
  {El-Showk}, \citenamefont {Paulos}, \citenamefont {Poland}, \citenamefont
  {Rychkov}, \citenamefont {Simmons-Duffin},\ and\ \citenamefont
  {Vichi}}]{elshowk2014conformal}%
  \BibitemOpen
  \bibfield  {author} {\bibinfo {author} {\bibfnamefont {S.}~\bibnamefont
  {El-Showk}}, \bibinfo {author} {\bibfnamefont {M.}~\bibnamefont {Paulos}},
  \bibinfo {author} {\bibfnamefont {D.}~\bibnamefont {Poland}}, \bibinfo
  {author} {\bibfnamefont {S.}~\bibnamefont {Rychkov}}, \bibinfo {author}
  {\bibfnamefont {D.}~\bibnamefont {Simmons-Duffin}},\ and\ \bibinfo {author}
  {\bibfnamefont {A.}~\bibnamefont {Vichi}},\ }\bibfield  {title} {\bibinfo
  {title} {Conformal field theories in fractional dimensions},\ }\href
  {https://doi.org/10.1103/PhysRevLett.112.141601} {\bibfield  {journal}
  {\bibinfo  {journal} {Phys. Rev. Lett.}\ }\textbf {\bibinfo {volume} {112}},\
  \bibinfo {pages} {141601} (\bibinfo {year} {2014})}\BibitemShut {NoStop}%
\bibitem [{\citenamefont {Sak}(1973)}]{Sak1973recursion}%
  \BibitemOpen
  \bibfield  {author} {\bibinfo {author} {\bibfnamefont {J.}~\bibnamefont
  {Sak}},\ }\bibfield  {title} {\bibinfo {title} {Recursion relations and fixed
  points for ferromagnets with long-range interactions},\ }\href
  {https://doi.org/10.1103/PhysRevB.8.281} {\bibfield  {journal} {\bibinfo
  {journal} {Phys. Rev. B}\ }\textbf {\bibinfo {volume} {8}},\ \bibinfo {pages}
  {281} (\bibinfo {year} {1973})}\BibitemShut {NoStop}%
\bibitem [{\citenamefont {Ba\~nos}\ \emph {et~al.}(2012)\citenamefont
  {Ba\~nos}, \citenamefont {Fernandez}, \citenamefont {Martin-Mayor},\ and\
  \citenamefont {Young}}]{Banos2012correspondence}%
  \BibitemOpen
  \bibfield  {author} {\bibinfo {author} {\bibfnamefont {R.~A.}\ \bibnamefont
  {Ba\~nos}}, \bibinfo {author} {\bibfnamefont {L.~A.}\ \bibnamefont
  {Fernandez}}, \bibinfo {author} {\bibfnamefont {V.}~\bibnamefont
  {Martin-Mayor}},\ and\ \bibinfo {author} {\bibfnamefont {A.~P.}\ \bibnamefont
  {Young}},\ }\bibfield  {title} {\bibinfo {title} {Correspondence between
  long-range and short-range spin glasses},\ }\href
  {https://doi.org/10.1103/PhysRevB.86.134416} {\bibfield  {journal} {\bibinfo
  {journal} {Phys. Rev. B}\ }\textbf {\bibinfo {volume} {86}},\ \bibinfo
  {pages} {134416} (\bibinfo {year} {2012})}\BibitemShut {NoStop}%
\bibitem [{\citenamefont {Dziarmaga}(2010)}]{Dziarmaga2010dynamics}%
  \BibitemOpen
  \bibfield  {author} {\bibinfo {author} {\bibfnamefont {J.}~\bibnamefont
  {Dziarmaga}},\ }\bibfield  {title} {\bibinfo {title} {Dynamics of a quantum
  phase transition and relaxation to a steady state},\ }\href
  {https://doi.org/10.1080/00018732.2010.514702} {\bibfield  {journal}
  {\bibinfo  {journal} {Advances in Physics}\ }\textbf {\bibinfo {volume}
  {59}},\ \bibinfo {pages} {1063} (\bibinfo {year} {2010})},\ \Eprint
  {https://arxiv.org/abs/https://doi.org/10.1080/00018732.2010.514702}
  {https://doi.org/10.1080/00018732.2010.514702} \BibitemShut {NoStop}%
\end{thebibliography}
%
\end{document}